# On the precoder design of flat fading MIMO systems equipped with MMSE receivers: a large system approach.

C. Artigue Freescale Semiconductor

134 Avenue du General Eisenhower - B.P. 29

31023 Toulouse Cedex 1, France

P. Loubaton Universite de Paris-Est

IGM LabInfo, UMR-CNRS 8049

5, Bd Descartes, Champs sur Marne

77454 Marne la Vallee Cedex 2, France

### Abstract

This paper is devoted to the design of precoders maximizing the ergodic mutual information (EMI) of bi-correlated flat fading MIMO systems equiped with MMSE receivers. The channel state information and the second order statistics of the channel are assumed available at the receiver side and at the transmitter side respectively. As the direct maximization of the EMI needs the use of non attractive algorithms, it is proposed to optimize an approximation of the EMI, introduced recently, obtained when the number of transmit and receive antennas $t$ and $r$ converge to $\infty$ at the same rate. It is established that the relative error between the actual EMI and its approximation is a $O(\frac{1}{t^2})$ term. It is shown that the left singular eigenvectors of the optimum precoder coincide with the eigenvectors of the transmit covariance matrix, and its singular values are solution of a certain maximization problem. Numerical experiments show that the mutual information provided by this precoder is close from what is obtained by maximizing the true EMI, but that the algorithm maximizing the approximation is much less computationally intensive.

## I. INTRODUCTION

It is now well established that using multiple transmit and receive antennas potentially allows to increase the Shannon capacity of digital communications systems. Since the seminal work of Teletar ([17]), the ergodic Shannon capacity of block fading MIMO systems has been studied extensively and important questions related to the design of optimal precoding schemes have been addressed. Considering that the Channel State Information (CSI) is available at receiver side while the transmitter is only aware of its second order statistical properties, many authors have studied the impact of antenna correlation on the capacity of MIMO systems communicating through flat fading channel ([7], [10]) and frequency selective channel ([12]).

The ergodic Shannon capacity is certainly a valuable figure of merit if the MIMO system under consideration is equipped with a maximum likelihood decoder. As the practical implementation of this decoder requires a high computational cost, it is also useful to study potential performance of MIMO systems equiped with the MMSE receiver. The corresponding (Gaussian) ergodic mutual information (EMI), denoted $I_{mmse}$ in the following, is defined as the sum over the transmit antennas of the terms $\mathbb{E}(\log(1+\beta_j))$, where $\beta_j$ represents the output MMSE SINR associated to the stream sent by antenna $j$. The design of precoders maximizing $I_{mmse}$ is of course an important issue because the optimum value of $I_{mmse}$ represents the maximum rate that can be transmitted reliably when the MIMO system uses the MMSE receiver. This optimization problem has been extensively studied in the past, mainly if the CSI is available at the both the receiver and the transmitter (see e.g. [15]). It is however often unrealistic to assume the CSI available at the transmitter side in the context of mobile systems.

In the present paper, we consider a flat fading MIMO channel with separable correlation structure (Kronecker model). We assume that the channel matrix is known at the receiver side, but that only its transmit and receive covariance matrices are available at the transmitter side. We address the problem of designing precoders that maximize $I_{mmse}$. The expression of $I_{mmse}$ is rather complicated and thus difficult to maximize w.r.t. the precoding matrix. In particular, it seems difficult to establish that the left eigenvectors of an optimal precoding matrix coincide with the eigenvectors of the transmit correlation matrix as in the context of the evaluation of the Shannon ergodic capacity (see e.g. ([7]). Therefore, it is necessary to evaluate numerically both the singular values and the singular vectors of optimum precoding matrices, or equivalently to solve a $t^2$ dimensional optimization problem. Steepest descent algorithms require the use intensive Monte Carlo simulation technics in order to evaluate the gradient and/or the Hessian of the cost function (see e.g. [20] in the context of the evaluation of the Shannon capacity of correlated Rician channels). Moreover, the convergence of these algorithms is not guaranteed because $I_{mmse}$ is in general not concave. As in previous contributions addressing the behaviour of the Shannon capacity of MIMO systems ([21], [4], [16]), we propose to replace the maximization of $I_{mmse}$ by the maximization of an approximation obtained in the asymptotic regime $t \to \infty$, $r \to \infty$, $\frac{t}{r} \to c$, $c \in (0, \infty)$.

Large system approximation of $I_{mmse}$ was previously considered in the context of CDMA systems with i.i.d. spreading codes (see e.g. [18] and the references herein), which, in the downlink, are formally equivalent to a subclass of the MIMO systems considered in this paper when the spreading codes are Gaussian. The specific case of MIMO systems has also been considered (see e.g. [3], [21]). It was shown that the SINRs $(\beta_j)_{j=1,\ldots,t}$ converge towards deterministic terms depending on the transmit and receive covariance matrices (or their equivalent in the context of downlink CDMA systems). These results provide an obvious large system approximation $\hat{I}_{mmse}$ of $I_{mmse}$.

In this paper, we establish that the large system approximation $\hat{I}_{mmse}$ provides a $O(\frac{1}{t})$ relative error. This is a rather poor convergence rate compared to the large system approximations of the Shannon capacity whose relative errors are $O(\frac{1}{t^2})$ ([11], [6]). We therefore propose to use an improved large system approximation, denoted $\overline{I}_{mmse}$, first introduced in [9] in the case of independent identically distributed (i.i.d.) MIMO channels, and then generalized independently in the conference papers [1] and [13]. The derivations of [13] are based on the replica method, a useful and powerfull trick whose mathematical relevance has not yet been established in the present context, and thus differ from the large random matrix approach sketched in [1]. We show that the relative error associated to $\overline{I}_{mmse}$ is a $O(\frac{1}{t^2})$ term, thus improving the predictions of [1] ($O(\frac{1}{t^{3/2}})$ [1]) and [13] ($o(\frac{1}{t})$). The method we use to study the accuracy of $\overline{I}_{mmse}$ differs from [9] whose approach is somewhat similar to [8], a paper devoted to the asymptotic study of the SINRs $(\beta_j)_{j=1,\ldots,t}$. The transmit covariance matrices of the MIMO channels of [8] are diagonal. This assumption simplifies the analysis so that the approach of [9], [8] cannot be generalized to the case of general transmit covariance matrices. Next, we address the maximization of $\overline{I}_{mmse}$ w.r.t. the precoding matrix. We establish that the left singular vectors of an optimum precoder are the eigenvectors of the transmit covariance matrix and that its right eigenvectors matrix is equal to $\mathbf{I}_t$. The evaluation of a precoding matrix thus reduces to the evaluation of its singular values, a $t$-dimensional optimization problem. In general, the optimum singular values have no closed form expression. In order to get more insights on the optimum precoders, we consider the case of an uncorrelated MIMO channel for which it is possible to obtain in closed form the precoders which optimize the approximation $\hat{I}_{mmse}$. We show that the optimum precoders are the diagonal matrices whose entries are either 0, either all coincide with $\frac{t}{s}$ where $s$ is the number of non zero entries which depend on the signal to noise ratio. Therefore, the optimum transmission strategy coincides with an antenna selection scheme. Although it is not proved that the above strategy maximizes $I_{mmse}$, this result shows that, at least if $t$ is large enough, antenna selection may provide higher mutual informations $I_{mmse}$ than a uniform power allocation. The situation differs from what was shown initially by Telatar ([17]) in the context of the study of the Shannon ergodic capacity of i.i.d. channels: the Shannon capacity achieving covariance matrix coincides with $\mathbf{I}_t$. We also remark that our result establishes formally that $\hat{I}_{mmse}$ is in general not a concave function of the precoding matrix, and infer from this that $I_{mmse}$ is not concave as well. We finally consider the case of an arbitrary bicorrelated MIMO channel, and propose to evaluate the singular values of an optimum precoder using a classical gradient algorithm. Numerical results show that the precoding matrices evaluated by this algorithm provide nearly the same mutual informations as direct approaches maximizing $I_{mmse}$ while being computationally more attractive.

---

[1] The authors wish to thank Aris Moustakas for suggesting that the rate $O(\frac{1}{t^{3/2}})$ was probably pessimistic

This paper is organized as follows. Section II is devoted to presentation of the problem and to the underlying assumptions. In section III, we present the large system approximations $\hat{I}_{mmse}$ and $\overline{I}_{mmse}$ of $I_{mmse}$ and analyse their accuracies. Section IV studies the structure of the optimum precoders, and Section V addresses the optimization of $\overline{I}_{mmse}$.

**General Notations** In this paper, the notations $y$, $\mathbf{x}$, $\mathbf{M}$ stand for scalars, vectors and matrices, respectively. As usual, $\|\mathbf{x}\|$ represents the Euclidian norm of vector $\mathbf{x}$ and $\|\mathbf{M}\|$ stands for the spectral norm of matrix $\mathbf{M}$. The superscripts $(.)^T$ and $(.)^H$ represent respectively the transpose and transpose conjugate. The trace of $\mathbf{M}$ is denoted by $\mathrm{Tr}(\mathbf{M})$. The mathematical expectation operator is denoted by $\mathbb{E}(\cdot)$. The symbols $\Re$ and $\Im$ denote respectively the real and imaginary parts of a given complex number. If $x$ is a possibly complex-valued random variable, $\mathrm{Var}(x) = \mathbb{E}|x|^2 - |\mathbb{E}(x)|^2$ represents the variance of $x$.

All along this paper, $t$ and $r$ stand for the number of transmit and receive antennas. Certain quantities will be studied in the asymptotic regime $t \to \infty$, $r \to \infty$ in such a way that $\frac{t}{r} \to c \in (0, \infty)$. In order to simplify the notations, $t \to \infty$ should be understood from now on as $t \to \infty$, $r \to \infty$ and $\frac{t}{r} \to c \in (0, \infty)$. A vector $\mathbf{x}_t$ and a matrix $\mathbf{M}_t$ whose size depend on $t$ are said to be uniformly bounded if $\sup_t \|\mathbf{x}_t\| < \infty$ and $\sup_t \|\mathbf{M}_t\| < \infty$.

Several variables used throughout this paper depend on various parameters, e.g. the number of antennas, the noise level, etc. In order to simplify the notations, we do not mention all these dependencies.

 

Notation $C$ will denote a generic strictly positive constant whose main feature is not to depend on $t$. The value of $C$ might change from one line to another.

## II. PROBLEM STATEMENT.

We consider a MIMO system equipped with $r$ receive antennas and $t$ transmit antennas. The MIMO channel matrix $\mathbf{H}$ is supposed to be a Gaussian random matrix defined by

$$\mathbf{H} = \frac{1}{\sqrt{t}} \mathbf{C}_R^{1/2} \mathbf{H}_{iid} \mathbf{C}_T^{1/2} \tag{1}$$

where $\mathbf{H}_{iid}$ is a $r \times t$ matrix whose entries are independent and identically distributed (i.i.d.) complex circular Gaussian random variables $\mathcal{CN}(0, 1)$, i.e. $H_{iid,ij} = \Re H_{iid,ij} + \mathbf{i}\Im H_{iid,ij}$ where $\Re H_{iid,ij}$ and $\Im H_{iid,ij}$ are independent centered real Gaussian random variables with variance $\frac{1}{2}$. Matrices $\mathbf{C}_T$ and $\mathbf{C}_R$ are positive definite matrices modeling respectively the impact of correlation between transmitting and receiving antennas. We assume that $\frac{1}{t}\text{Trace}\,(\mathbf{C}_T) = 1$ and $\frac{1}{r}\text{Trace}\,(\mathbf{C}_R) = 1$. This assumption implies that $\frac{1}{r}\mathbb{E}(\text{Tr}(\mathbf{H}\mathbf{H}^H)) = 1$.

Each transmit antenna $j$ sends a sequence $(x_j(n))_{n \in \mathbb{Z}}$ defined by

$$\mathbf{x}(n) = (x_1(n), \ldots, x_t(n))^T = \mathbf{K}\mathbf{s}(n) = \mathbf{K}(s_1(n), \ldots, s_t(n))^T$$

where the $\big((s_j(n))_{n \in \mathbf{Z}}\big)_{j=1,\ldots,t}$ are assumed to be unit variance mutually independent i.i.d. sequences. $\mathbf{K}$ represents a precoding matrix satisfying $\frac{1}{t}\text{Tr}(\mathbf{K}\mathbf{K}^H) \leq 1$.

The corresponding $r$-variate discrete-time received signal $(\mathbf{y}(n))_{n \in \mathbf{Z}}$ is given by

$$\mathbf{y}(n) = \mathbf{H}\mathbf{K}\mathbf{s}(n) + \mathbf{n}(n) \tag{2}$$

where $\mathbf{n}$ is a white Gaussian noise with covariance matrix $\mathbf{E}\,(\mathbf{n}(n)\mathbf{n}(n)^H) = \sigma^2 \mathbf{I}_r$.

In this paper, we evaluate the potential performance of the MIMO system (2) when the receiver is equiped with the MMSE receiver. In other words, each symbol sequence $s_j$ is estimated by the Wiener filter prior to decoding, i.e. $s_j(n)$ is estimated by

$$\hat{s}_j(n) = \mathbf{k}_j^H \mathbf{H}^H \left(\mathbf{H}\mathbf{K}\mathbf{K}^H\mathbf{H}^H + \sigma^2 \mathbf{I}_r\right)^{-1} \mathbf{y}(n)$$

where $\mathbf{k}_j$ represents the column $j$ of $\mathbf{K}$. In the following, we denote by $\mathbf{Q}_T(\mathbf{K})$ the matrix

$$\mathbf{Q}_T(\mathbf{K}) = \left(\mathbf{K}^H \mathbf{H}^H \mathbf{H}\mathbf{K} + \sigma^2 \mathbf{I}_t\right)^{-1} \tag{3}$$

It is standard that the SINR $\beta_j$ provided by this linear receiver is given by ([19])

$$\beta_j(\mathbf{K}) = \frac{1}{\sigma^2(\mathbf{Q}_T(\mathbf{K}))_{j,j}} - 1 \tag{4}$$

 

The ergodic mutual information $I_{mmse}(\mathbf{K})$ of the MIMO system under consideration is thus equal to

$$I_{mmse}(\mathbf{K}) = \mathbb{E}\left[\sum_{j=1}^{t}\log\left(1 + \beta_j(\mathbf{K})\right)\right] = -\mathbb{E}\left[\sum_{j=1}^{t}\log\left(\sigma^2\mathbf{Q}_T(\mathbf{K})\right)_{j,j}\right] \qquad (5)$$

where the mathematical expectation is over the probability distribution of random matrix $\mathbf{H}$. In order to maximize $I_{mmse}(\mathbf{K})$ over the set $\frac{1}{t}\mathrm{Tr}(\mathbf{K}\mathbf{K}^H) \leq 1$, it is necessary to use numerical technics based on stepeest descent algorithms. As the gradient and the Hessian of $I_{mmse}$ have no simple expression, they have to be evaluated using intensive Monte Carlo simulations (see e.g. [20]). Moreover, to our best knowledge, the singular vectors of an optimum matrix have no closed form expression. Therefore, the dimension of the optimization problem cannot be reduced from $t^2$ to $t$ as in the context of the evaluation of the capacity achieving covariance matrix ([7]).

## III. Derivation of the large system approximation of $I_{mmse}$.

In this section, we introduce the large system approximation presented in [1] and [13], and improve the results stated without proof in [1] concerning its accuracy. Our approach is based on Gaussian large random matrix technics initiated by Pastur ([14]). Pastur's approach was used in [6] in order to establish the asymptotic Gaussianity of the traditional mutual information of bicorrelated MIMO channels.

We study in this section the asymptotic behaviour of $I_{mmse}$ in the case where the precoding matrix $\mathbf{K}$ is reduced to $\mathbf{K} = \mathbf{I}_t$ to simplify the notations. In order to deduce the results in the case $\mathbf{K} \neq \mathbf{I}_t$, we remark that channel matrix $\mathbf{H}\mathbf{K}$ can be interpreted as a bi-correlated MIMO channel with transmit and receive covariance matrices $\mathbf{K}^H\mathbf{C}_T\mathbf{K}$ and $\mathbf{C}_R$ respectively. We will therefore replace matrix $\mathbf{C}_T$ by matrix $\mathbf{K}^H\mathbf{C}_T\mathbf{K}$. $I_{mmse}(\mathbf{I})$ and $\mathbf{Q}_T(\mathbf{I})$ are denoted $I_{mmse}$ and $\mathbf{Q}_T$ in the remainder of this section.

We first explain the differences between our analysis and the contributions [9] and [8] . We recall that [9] adresses the i.i.d. case while [8] assumes that matrix $\mathbf{C}_T = \mathrm{Diag}(c_{T,1}, \ldots, c_{T,t})$ is diagonal. In this last context, the SINR $\beta_j$ can also be written as

$$\beta_j = \frac{c_{T,j}}{t}\mathbf{h}_{iid,j}^H\left(\mathbf{C}_R^{1/2}\mathbf{H}_{iid}^{(j)}\frac{\mathbf{C}_T^{(j)}}{t}\mathbf{H}_{iid}^{(j)H}\mathbf{C}_R^{1/2}\right)^{-1}\mathbf{h}_{iid,j} = \frac{c_{T,j}}{t}\mathbf{h}_{iid,j}^H\,\mathbf{Q}_R^{(j)}\,\mathbf{h}_{iid,j} \qquad (6)$$

where $\mathbf{h}_{iid}^{(j)}$ represents the column $j$ of $\mathbf{H}_{iid}$, matrix $\mathbf{H}_{iid}^{(j)}$ is obtained from $\mathbf{H}_{iid}$ by deleting column $j$, and $\mathbf{C}_T^{(j)}$ represents the $(t-1) \times (t-1)$ diagonal matrix obtained by deleting column and row $j$ from $\mathbf{C}_T$. The approaches of [9] and [8] rely on the key observation that vector $\mathbf{h}_{iid,j}$ is independent from the matrix $\mathbf{Q}_R^{(j)}$. This allows to study the behaviour of $\beta_j$ using important results concerning the behaviour of random quadratic forms. If matrix $\mathbf{C}_T$ is non diagonal, $\beta_j$ has not the same structure than in (6): vector $\mathbf{h}_{iid}^{(j)}$ and matrices $\mathbf{Q}_R^{(j)}$ are replaced by non independent terms, and the approach of [9] and [8] cannot be used. Our approach does not study $\beta_j$ directly, but rather the diagonal entries of matrix $\sigma^2 \mathbf{Q}_T$ whose asymptotic behaviour can be evaluated for general transmit covariance matrices $\mathbf{C}_T$.

The study of the accuracy of the approximation is essentially based on the study of a virtual channel obtained from $\mathbf{H}$ after unitary transformations. We consider the eigenvalue/ eigenvector decompositions of covariance matrices $\mathbf{C}_T$ and $\mathbf{C}_R$:

$$\mathbf{C}_T = \mathbf{U}\mathbf{D}\mathbf{U}^H, \quad \mathbf{C}_R = \tilde{\mathbf{U}}\tilde{\mathbf{D}}\tilde{\mathbf{U}}^H \tag{7}$$

where the diagonal entries $(d_j)_{j=1,\ldots,t}$ and $(\tilde{d}_i)_{i=1,\ldots,r}$ of $\mathbf{D}$ and $\tilde{\mathbf{D}}$ are arranged in the decreasing order. Then, we define the random $t \times r$ matrix $\mathbf{Y}$ by

$$\mathbf{Y}^H = \tilde{\mathbf{U}}^H \mathbf{H} \mathbf{U} \tag{8}$$

$\mathbf{Y}$ can be written as

$$\mathbf{Y} = \frac{1}{\sqrt{t}} \mathbf{D}^{1/2} \mathbf{X} \tilde{\mathbf{D}}^{1/2} \tag{9}$$

where $\mathbf{X}$ represents the $t \times r$ matrix $\mathbf{X} = \mathbf{U}^H \mathbf{H}_{iid}^H \tilde{\mathbf{U}}$. As $\mathbf{U}$ and $\tilde{\mathbf{U}}$ are unitary, matrix $\mathbf{X}$ is an i.i.d. complex Gaussian matrix such that $\mathbb{E}|X_{i,j}|^2 = 1$. In the following, we denote by $\mathbf{Q}$ the matrix defined by

$$\mathbf{Q} = \left(\mathbf{Y}\mathbf{Y}^H + \sigma^2 \mathbf{I}\right)^{-1} \tag{10}$$

The study of $I_{mmse}$ when $t \to \infty$ is based on the asymptotic properties of the diagonal entries of matrix $\mathbf{Q}_T$. We remark that

$$\mathbf{Q}_T = \mathbf{U}\mathbf{Q}\mathbf{U}^H \tag{11}$$

and evaluate the asymptotic behaviour of $\mathbf{u}\mathbf{Q}\mathbf{u}^H$ where $\mathbf{u} = (u_1, \ldots, u_t)$ is a unit norm deterministic row vector. We use in the following certain results of [6]. We however note that in [6], matrix $\mathbf{Q}$ is replaced by matrix $(\mathbf{I} + \sigma^2 \mathbf{Y}\mathbf{Y}^H)^{-1}$. Therefore, the statements of [6] have to be adapted. In the sequel, we denote by $\delta$ and $\tilde{\delta}$ the unique strictly positive solutions of the system

$$\begin{array}{rcl} \delta &=& \frac{1}{t}\mathrm{Tr}\left[\mathbf{D}\left(\sigma^2(\mathbf{I} + \tilde{\delta}\mathbf{D})\right)\right]^{-1} \\ \tilde{\delta} &=& \frac{1}{t}\mathrm{Tr}\left[\tilde{\mathbf{D}}\left(\sigma^2(\mathbf{I} + \delta\tilde{\mathbf{D}})\right)\right]^{-1} \end{array} \tag{12}$$

The existence and the uniqueness of the solution has been established in Proposition 1 of [6]. We denote by $\mathbf{T}$ and $\tilde{\mathbf{T}}$ the diagonal matrices

$$\begin{array}{rcl} \mathbf{T} &=& \left[\sigma^2(\mathbf{I} + \tilde{\delta}\mathbf{D})\right]^{-1} \\ \tilde{\mathbf{T}} &=& \left[\sigma^2(\mathbf{I} + \delta\tilde{\mathbf{D}})\right]^{-1} \end{array} \tag{13}$$

and gather in the following proposition certain useful results of [6].

*Proposition 1:* Assume that matrices $\mathbf{D}$ and $\tilde{\mathbf{D}}$ satisfy the following conditions:

$$\sup_t \|\mathbf{D}\| \leq d_{max} < \infty \quad , \quad \inf_t \tfrac{1}{t}\mathrm{Tr}\mathbf{D} > 0$$
$$\sup_t \|\tilde{\mathbf{D}}\| \leq \tilde{d}_{max} < \infty \quad , \quad \inf_t \tfrac{1}{t}\mathrm{Tr}\tilde{\mathbf{D}} > 0 \tag{14}$$

Then, the following results hold true:

- For each uniformly bounded deterministic matrix $\mathbf{M}$ [2]

$$\mathrm{Var}\left(\tfrac{1}{t}\mathrm{Tr}(\mathbf{MQ})\right) = O(\tfrac{1}{t^2})$$
$$\mathbb{E}\left(\mathrm{Tr}\left(\mathbf{M}(\mathbf{Q}-\mathbf{T})\right)\right) = O(\tfrac{1}{t}) \tag{15}$$

- $\gamma$ and $\tilde{\gamma}$ defined by

$$\gamma = \frac{1}{t}\mathrm{Tr}\left(\mathbf{D}^2\mathbf{T}^2\right), \; \tilde{\gamma} = \frac{1}{t}\mathrm{Tr}\left(\tilde{\mathbf{D}}^2\tilde{\mathbf{T}}^2\right), \tag{16}$$

    satisfy

$$\inf_t \left(1 - \sigma^4\gamma\tilde{\gamma}\right) > 0 \tag{17}$$

We assume from now on that the matrices $\mathbf{D}$ and $\tilde{\mathbf{D}}$ satisfy (14). We are now in position to state the main results of this section. We begin by the following proposition.

*Proposition 2:*

$$\sup_{\mathbf{u},\|\mathbf{u}\|=1} \left|\mathbb{E}\left(\mathbf{u}(\mathbf{Q}-\mathbf{T})\mathbf{u}^H\right)\right| = O(\frac{1}{t^{3/2}}) \tag{18}$$

and

$$\sup_{\mathbf{u},\|\mathbf{u}\|=1} \left|\mathbb{E}\left(\mathbf{u}(\mathbf{Q}-\mathbb{E}(\mathbf{Q}))\mathbf{u}^H\right)^3\right| = O(\frac{1}{t^2}) \tag{19}$$

Moreover,

$$\sup_{\mathbf{u},\|\mathbf{u}\|=1} \left|\mathrm{Var}\left(\mathbf{u}\mathbf{Q}\mathbf{u}^H\right) - \frac{1}{t}\frac{\sigma^4\tilde{\gamma}}{1-\sigma^4\gamma\tilde{\gamma}}\left(\mathbf{u}\mathbf{T}^2\mathbf{D}\mathbf{u}^H\right)^2\right| = O(\frac{1}{t^{3/2}}) \tag{20}$$

Finally, if we denote by $(\mathbf{v}_k)_{k=1,\ldots,t}$ the row vectors of any unitary matrix $\mathbf{V}$, and if $(\kappa_j)_{j=1,\ldots,t}$ denote positive numbers such that $\sup_j \kappa_j < C$, we have

$$\sum_{k=1}^t \left[\mathrm{Var}\left(\kappa_k\mathbf{v}_k\mathbf{Q}\mathbf{v}_k^H\right) - \frac{1}{t}\frac{\sigma^4\tilde{\gamma}}{1-\sigma^4\gamma\tilde{\gamma}}\left(\kappa_k\mathbf{v}_k\mathbf{T}^2\mathbf{D}\mathbf{v}_k^H\right)^2\right] = O(\frac{1}{t}) \tag{21}$$

The proof is given in the Appendix. In order to introduce the large system approximation $\overline{I}_{mmse}$, we define matrices $\mathbf{T}_T$ and $\mathbf{T}_R$ by

$$\mathbf{T}_T = \mathbf{U}\mathbf{T}\mathbf{U}^H = \left(\sigma^2(\mathbf{I}+\tilde{\delta}\mathbf{C}_T)\right)^{-1}, \; \mathbf{T}_R = \tilde{\mathbf{U}}\tilde{\mathbf{T}}\tilde{\mathbf{U}}^H = \left(\sigma^2(\mathbf{I}+\delta\mathbf{C}_R)\right)^{-1} \tag{22}$$

---

[2]In [6], matrix $\mathbf{M}$ is diagonal. The case of non diagonal matrices is addressed in [5] devoted to correlated Ricean channels.

We note that $(\delta, \tilde{\delta})$ and $(\gamma, \tilde{\gamma})$ can be expressed in terms of $\mathbf{T}_T$ and $\mathbf{T}_R$ as

$$
\begin{array}{rcl}
\delta & = & \frac{1}{t}\mathrm{Tr}\left[\mathbf{C}_T\left(\sigma^2(\mathbf{I}+\tilde{\delta}\mathbf{C}_T)\right)^{-1}\right] = \frac{1}{t}\mathrm{Tr}\left[\mathbf{C}_T\mathbf{T}_T\right] \\
\tilde{\delta} & = & \frac{1}{t}\mathrm{Tr}\left[\mathbf{C}_R\left(\sigma^2(\mathbf{I}+\delta\mathbf{C}_R)\right)^{-1}\right] = \frac{1}{t}\mathrm{Tr}\left[\mathbf{C}_R\mathbf{T}_R\right]
\end{array}
\tag{23}
$$

and

$$
\gamma = \frac{1}{t}\mathrm{Tr}(\mathbf{C}_T^2\mathbf{T}_T^2), \; \tilde{\gamma} = \frac{1}{t}\mathrm{Tr}(\mathbf{C}_R^2\mathbf{T}_R^2)
\tag{24}
$$

The following result holds.

*Theorem 1:* We define $\overline{I}_{mmse}$ by

$$
\overline{I}_{mmse} = -\sum_{j=1}^{t}\log(\sigma^2\mathbf{T}_{T,j,j}) + \frac{1}{2\tilde{\delta}^2}\frac{\tilde{\gamma}}{1-\sigma^4\gamma\tilde{\gamma}}\frac{1}{t}\sum_{j=1}^{t}\left(1-\frac{\left((\sigma^2\mathbf{T}_T)^2\right)_{j,j}}{\sigma^2\mathbf{T}_{T,j,j}}\right)^2
\tag{25}
$$

Then,

$$
I_{mmse} = \overline{I}_{mmse} + O(\frac{1}{t})
\tag{26}
$$

**Proof.** The proof is based on a second order expansion of $\log\sigma^2\mathbf{Q}_{T,j,j}$ around the point $\mathbb{E}(\mathbf{Q}_{T,j,j})$ . We define $\epsilon_j$ by

$$
\epsilon_j = \frac{\mathbf{Q}_{T,j,j} - \mathbb{E}(\mathbf{Q}_{T,j,j})}{\mathbb{E}(\mathbf{Q}_{T,j,j})}
$$

and write $\log\sigma^2\mathbf{Q}_{T,j,j}$ as

$$
\log\sigma^2\mathbf{Q}_{T,j,j} = \log\left(\sigma^2\mathbb{E}(\mathbf{Q}_{T,j,j})\right) + \log(1+\epsilon_j)
$$

We express $\log(1+\epsilon_j)$ as

$$
\log(1+\epsilon_j) = \epsilon_j - \frac{\epsilon_j^2}{2} + \frac{\epsilon_j^3}{3} + r_j
$$

As $\mathbb{E}(\epsilon_j) = 0$, $\mathbb{E}(\log\sigma^2\mathbf{Q}_{T,j,j})$ can be written as

$$
\mathbb{E}(\log\sigma^2\mathbf{Q}_{T,j,j}) = \log\left(\sigma^2\mathbb{E}(\mathbf{Q}_{T,j,j})\right) - \frac{1}{2}\mathbb{E}(\epsilon_j^2) + \frac{1}{3}\mathbb{E}(\epsilon_j^3) + \mathbb{E}(r_j)
\tag{27}
$$

In order to be able to use Proposition 2, we have to study the behaviour $\frac{1}{\mathbb{E}(\mathbf{Q}_{T,j,j})}$. We first remark that it exists a deterministic constant $C > 0$ such that

$$
\sup_{j,t}\frac{1}{\mathbb{E}(\mathbf{Q}_{T,j,j})} \leq C
\tag{28}
$$

Indeed, $\frac{1}{\mathbb{E}(\mathbf{Q}_{T,j,j})} \leq \mathbb{E}\left(\frac{1}{\mathbf{Q}_{T,j,j}}\right) = \sigma^2\mathbb{E}(1+\beta_j)$ by the Jensen inequality. We denote by $\mathbf{h}_j$ the column $j$ of matrix $\mathbf{H}$. The SINR $\beta_j$ provided by the MMSE receiver is upperbounded by the match filter bound, i.e. $\beta_j \leq \frac{\|\mathbf{h}_j\|^2}{\sigma^2}$. As it is clear that $\sup_{j,t}\mathbb{E}(\|\mathbf{h}_j\|^2) \leq C$, we get (28).

We use (21) with $\kappa_j = \frac{1}{\mathbb{E}(\mathbf{Q}_{T,j,j})}$ and when the unitary matrix $\mathbf{V}$ coincides with matrix $\mathbf{U}$. We obtain immediately that

$$\sum_{j=1}^{t} \mathbb{E}(\epsilon_j^2) - \frac{1}{t}\frac{\sigma^4\tilde{\gamma}}{1-\sigma^4\gamma\tilde{\gamma}}\left(\frac{\mathbf{u}_j\mathbf{T}^2\mathbf{D}\mathbf{u}_j^H}{\mathbb{E}(\mathbf{u}_j\mathbf{Q}\mathbf{u}_j^H)}\right)^2 = O(\frac{1}{t}) \tag{29}$$

We now establish that

$$\sup_j \left|\frac{1}{\mathbb{E}(\mathbf{Q}_{T,j,j})} - \frac{1}{\mathbf{T}_{T,j,j}}\right| \le \frac{C}{t^{3/2}} \tag{30}$$

For this, we first notice that

$$\frac{1}{\mathbf{T}_{T,j,j}} \le C \tag{31}$$

Indeed,

$$\frac{1}{\mathbf{T}_{T,j,j}} \le (\mathbf{T}_T)_{j,j} = \sigma^2\left(1 + \tilde{\delta}\mathbf{C}_{T,j,j}\right)$$

The conclusion follows $\tilde{\delta} \le \frac{1}{\sigma^2}\frac{1}{t}\text{Tr}(\tilde{\mathbf{D}}) \le \frac{r\tilde{d}_{max}}{t\sigma^2}$. (30) follows directly from

$$\frac{1}{\mathbb{E}(\mathbf{Q}_{T,j,j})} - \frac{1}{\mathbf{T}_{T,j,j}} = \frac{\mathbf{T}_{T,j,j} - \mathbb{E}(\mathbf{Q}_{T,j,j})}{\mathbf{T}_{T,j,j}\mathbb{E}(\mathbf{Q}_{T,j,j})}$$

and from (18).

(29) and (30) imply that

$$\sum_{j=1}^{t} \mathbb{E}(\epsilon_j^2) - \frac{1}{t}\frac{\sigma^4\tilde{\gamma}}{1-\sigma^4\gamma\tilde{\gamma}}\left(\frac{\mathbf{u}_j\mathbf{T}^2\mathbf{D}\mathbf{u}_j^H}{\mathbf{u}_j\mathbf{T}\mathbf{u}_j^H}\right)^2 = O(\frac{1}{t}) \tag{32}$$

Moreover, (19) and (28) lead to

$$\sup_j \mathbb{E}(\epsilon_j^3) \le \frac{C}{t^2} \tag{33}$$

In order to evaluate the influence of $r_j$, we give the following lemma, proved in the appendix.

*Lemma 1:*

$$\sup_j |\mathbb{E}(r_j)| = O(\frac{1}{t^2}) \tag{34}$$

(32) and (34) imply that

$$I_{mmse} = -\sum_{j=1}^{t}\log\mathbb{E}(\sigma^2\mathbf{Q}_{T,j,j}) + \frac{1}{2t}\frac{\sigma^4\tilde{\gamma}}{1-\sigma^4\gamma\tilde{\gamma}}\sum_{j=1}^{t}\left(\frac{\mathbf{u}_j\mathbf{T}^2\mathbf{D}\mathbf{u}_j^H}{\mathbf{u}_j\mathbf{T}\mathbf{u}_j^H}\right)^2 + O(\frac{1}{t})$$

Straightforward manipulations show that

$$\sum_{j=1}^{t}\left(\frac{\mathbf{u}_j\mathbf{T}^2\mathbf{D}\mathbf{u}_j^H}{\mathbf{u}_j\mathbf{T}\mathbf{u}_j^H}\right)^2 = \frac{1}{\tilde{\delta}^2}\sum_{j=1}^{t}\left(1 - \frac{\left((\sigma^2\mathbf{T})^2\right)_{j,j}}{\sigma^2\mathbf{T}_{T,j,j}}\right)^2$$

In order to establish Theorem 1, it remains to prove that

$$\sum_{j=1}^{t}\log\mathbb{E}(\sigma^2\mathbf{Q}_{T,j,j}) = \sum_{j=1}^{t}\log\left(\sigma^2\mathbf{T}_{T,j,j}\right) + O(\frac{1}{t}) \tag{35}$$

We define $\overline{\epsilon}_j$ as

$$\overline{\epsilon}_j = \frac{\mathbf{T}_{T,j,j} - \mathbb{E}(\mathbf{Q}_{T,j,j})}{\mathbb{E}(\mathbf{Q}_{T,j,j})}$$

and remark that

$$\log\left(\sigma^2 \mathbf{T}_{T,j,j}\right) = \log \mathbb{E}(\sigma^2 \mathbf{Q}_{T,j,j}) + \log(1 + \overline{\epsilon}_j)$$

Using (28), we obtain that

$$|\overline{\epsilon}_j| \leq C \left|\mathbf{T}_{T,j,j} - \mathbb{E}(\mathbf{Q}_{T,j,j})\right| \tag{36}$$

(18) implies that $\sup_j |\mathbf{T}_{T,j,j} - \mathbb{E}(\mathbf{Q}_{T,j,j})| = O(\frac{1}{t^{3/2}})$. By (36), we get that $\sup_j |\overline{\epsilon}_j| = O(\frac{1}{t^{3/2}})$. For $t$ large enough, $|\overline{\epsilon}_j| < A < 1$ for each $j$. For these $t$, we can write $\log(1 + \overline{\epsilon}_j)$ as

$$\log(1 + \overline{\epsilon}_j) = \overline{\epsilon}_j + \overline{r}_j$$

where

$$\overline{r}_j = (\overline{\epsilon}_j)^2 \sum_{n=2}^{\infty} (-1)^{n-1} \frac{\overline{\epsilon}_j^{n-2}}{n}$$

By (18), it holds that $\sup_j \overline{\epsilon}_j^2 = O(\frac{1}{t^3})$. Therefore,

$$|\overline{r}_j| < (\overline{\epsilon}_j)^2 \sum_{n=2}^{\infty} \frac{A^{n-2}}{n} < \frac{C}{t^3}$$

Consequently,

$$\sum_{j=1}^{t} \log(1 + \overline{\epsilon}_j) = \sum_{j=1}^{t} \overline{\epsilon}_j + O(\frac{1}{t^2})$$

We finally remark that

$$\sum_{j=1}^{t} \overline{\epsilon}_j = \sum_{j=1}^{t} \kappa_j \mathbf{u}_j \mathbb{E}\left(\mathbf{Q} - \mathbf{T}\right) \mathbf{u}_j^H$$

where $\kappa_j = \frac{1}{\mathbb{E}(\mathbf{Q}_{T,j,j})}$. The second item of (15) can thus be used for matrix $\mathbf{M} = \sum_j \kappa_j \mathbf{u}_j^H \mathbf{u}_j$, thus showing that (35) holds. This completes the proof of (26).

We denote $\hat{I}_{mmse}$ the term defined by

$$\hat{I}_{mmse} = -\sum_{j=1}^{t} \log(\sigma^2 \mathbf{T}_{T,j,j}) \tag{37}$$

$\hat{I}_{mmse}$ corresponds to the obvious large system approximation of $I_{mmse}$ obtained by replacing, for each $j$, $(1 + \beta_j)$ by its "deterministic equivalent" $(\sigma^2 \mathbf{T}_{T,j,j})^{-1}$. Theorem 1 shows that the relative error provided by $\hat{I}_{mmse}$ is a $O(\frac{1}{t})$ term, while the relative error of $\overline{I}_{mmse}$ is a $O(\frac{1}{t^2})$ term.

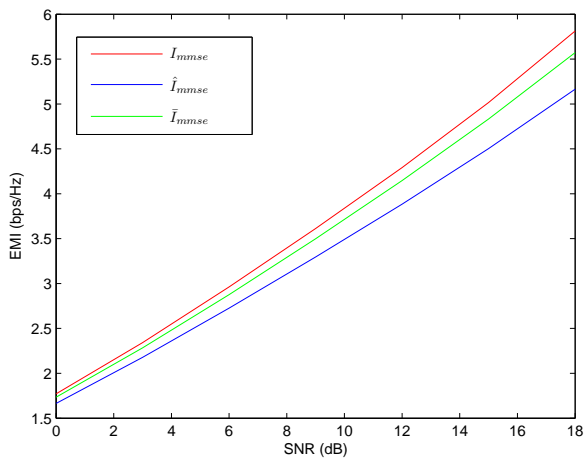

Fig. 1. Accuracy of the large system approximant

We now present some simulation experiments which demonstrate the accuracy of the approximation $\overline{I}_{mmse}$ for a realistic number of antennas. $\hat{I}_{mmse}$ is also represented. The transmit antennas correlation matrix $\mathbf{C}_T$ is generated according to the popular model proposed in [2], i.e.

$$\mathbf{C}_{T,k,l} = a\,e^{-i\pi(k-l)\cos\phi_T}\,e^{-\frac{1}{2}(\pi(k-l)\sin\phi_T\,\sigma_{\phi_T})^2} \tag{38}$$

where $a$ is a constant chosen in such a way that $\frac{1}{t}\mathrm{Tr}(\mathbf{C}_T) = 1$. $\phi_T$ and $\sigma_{\phi_T}$ can be interpreted as the mean angle of departure and the standard deviation of the angles of departure of a scatterer cluster respectively. We notice that if $\sigma_{\phi_T} \simeq 0$, then $\mathrm{Rank}(\mathbf{C}_T) \simeq 1$. We refer the reader to [2] for more details.

The receive antennas correlation matrix is generated similarly with different parameters $\phi_R$ and $\sigma_{\phi_R}$.

In Figure 1 we have represented $I_{mmse}, \hat{I}_{mmse}, \overline{I}_{mmse}$ versus the SNR for $r = t = 4$. Here, the various parameters are equal to $\phi_T = \pi/4, \sigma_{\phi_T} = 0.5, \phi_R = \pi/12, \sigma_{\phi_R} = 0.5$. We observe that $\hat{I}_{mmse}$ can be rather far from the true mutual information $I_{mmse}$ evaluated by Monte-Carlo simulation over 1000 channel realizations. Figure 2 represents the relative error between $I_{mmse}$ and $\hat{I}_{mmse}, \overline{I}_{mmse}$ respectively in terms of the mean angle of departure variance $\sigma_{\phi_T}^2$ for SNR = 0 dB and SNR = 6 dB when $\phi_T = \pi/4, \phi_R = \pi/12, \sigma_{\phi_R} = 0.4$ . Figures 1,2 show that approximation $\overline{I}_{mmse}$ provides significantly better results than $\hat{I}_{mmse}$.

The expression (25) is a large system approximation of $I_{mmse}(\mathbf{I})$. If the precoding matrix $\mathbf{K}$ is not equal to $\mathbf{I}$, the approximation of $I_{mmse}(\mathbf{K})$ is obtained by replacing matrix $\mathbf{C}_T$ by matrix $\mathbf{K}^H\mathbf{C}_T\mathbf{K}$. In the following we denote by $\delta(\mathbf{K}), \tilde{\delta}(\mathbf{K}), \mathbf{T}_T(\mathbf{K}), \mathbf{T}_R(\mathbf{K}), \gamma(\mathbf{K}), \tilde{\gamma}(\mathbf{K}), \overline{I}_{mmse}(\mathbf{K}), \hat{I}_{mmse}(\mathbf{K})$ the values of parameters $\delta, \tilde{\delta}, \mathbf{T}_T, \mathbf{T}_R, \gamma, \tilde{\gamma}, \overline{I}_{mmse}, \hat{I}_{mmse}$ when $\mathbf{C}_T$ is replaced by $\mathbf{K}^H \mathbf{C}_T \mathbf{K}$.

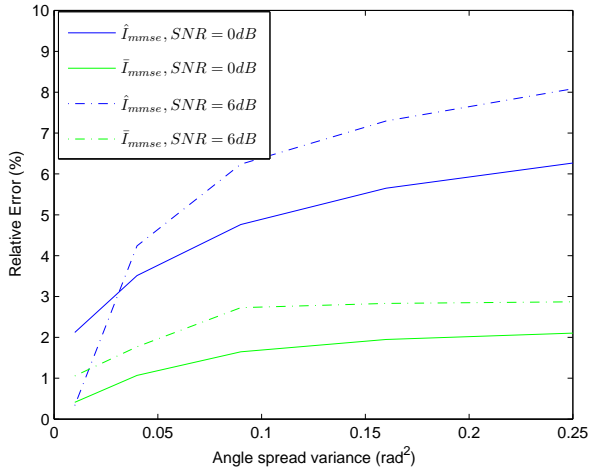

Fig. 2. Relative error

## IV. Structure of optimal precoders.

In this section, we study the problem of designing precoders maximizing function $\overline{I}_{mmse}(\mathbf{K})$ over the set $\mathcal{K}$ defined by

$$\mathcal{K} = \{\mathbf{K}, \frac{1}{t}\text{Tr}(\mathbf{K}\mathbf{K}^H) \leq 1\} \tag{39}$$

The main result of this section states that there is no restriction to look for optimal precoders of the form $\mathbf{K} = \mathbf{U}\mathbf{D}^{-1/2}\mathbf{\Lambda}^{1/2}$ where $\mathbf{\Lambda}$ is a diagonal matrix with positive elements. In order to establish this, we first derive the following intermediate result.

*Proposition 3:* Let $\mathbf{K}$ by an element of $\mathcal{K}$ and the eigenvalue/eigenvector decomposition of matrix $\mathbf{K}^H \mathbf{C}_T \mathbf{K}$

$$\mathbf{K}^H \mathbf{C}_T \mathbf{K} = \mathbf{W}\mathbf{\Lambda}\mathbf{W}^H$$

Then, matrix $\mathbf{K}_d = \mathbf{K}\mathbf{W}$ belongs to $\mathcal{K}$ and satisfies

$$\overline{I}_{mmse}(\mathbf{K}) \leq \overline{I}_{mmse}(\mathbf{K}_d) \tag{40}$$

**Proof.** It is obvious that $\mathbf{K}_d \in \mathcal{K}$. In order to establish (40), we denote by $\overline{J}_{mmse}(\mathbf{K})$ the term

$$\overline{J}_{mmse}(\mathbf{K}) = \frac{1}{2\tilde{\delta}(\mathbf{K})^2} \frac{\tilde{\gamma}(\mathbf{K})}{1 - \sigma^4 \gamma(\mathbf{K})\tilde{\gamma}(\mathbf{K})} \frac{1}{t} \sum_{j=1}^{t} \left(1 - \frac{(\sigma^2 \mathbf{T}_T(\mathbf{K}))^2)_{j,j}}{\sigma^2 \mathbf{T}_{T,j,j}(\mathbf{K})}\right)^2 \tag{41}$$

and remark that $\overline{I}_{mmse}(\mathbf{K}) = \hat{I}_{mmse}(\mathbf{K}) + \overline{J}_{mmse}(\mathbf{K})$. We prove in the following that $\hat{I}_{mmse}(\mathbf{K}) \leq \hat{I}_{mmse}(\mathbf{K}_d)$ and $\overline{J}_{mmse}(\mathbf{K}) \leq \overline{J}_{mmse}(\mathbf{K}_d)$.

We first remark that $\mathbf{K}_d^H \mathbf{C}_T \mathbf{K}_d$ is the diagonal matrix $\mathbf{\Lambda}$. Therefore, by (22) matrix $\sigma^2 \mathbf{T}_T(\mathbf{K}_d)$ is also diagonal, and is given by

$$\sigma^2 \mathbf{T}_T(\mathbf{K}_d) = \left( \mathbf{I} + \tilde{\delta}(\mathbf{K}_d)\mathbf{\Lambda} \right)^{-1}$$

Moreover,

$$\sigma^2 \mathbf{T}_T(\mathbf{K}) = \mathbf{W} \sigma^2 \mathbf{T}_T(\mathbf{K}_d) \mathbf{W}^H \tag{42}$$

We claim that $(\delta(\mathbf{K}), \tilde{\delta}(\mathbf{K})) = (\delta(\mathbf{K}_d), \tilde{\delta}(\mathbf{K}_d))$. To check this, we recall that $(\delta(\mathbf{K}), \tilde{\delta}(\mathbf{K}))$ are defined as the unique positive solutions of the system

$$\delta(\mathbf{K}) = \frac{1}{t} \mathrm{Tr} \mathbf{K}^H \mathbf{C}_T \mathbf{K} \left[ \sigma^2 (\mathbf{I} + \tilde{\delta}(\mathbf{K}) \mathbf{K}^H \mathbf{C}_T \mathbf{K}) \right]^{-1}$$

$$\tilde{\delta}(\mathbf{K}) = \frac{1}{t} \mathrm{Tr} \mathbf{C}_R \left[ \sigma^2 (\mathbf{I} + \delta(\mathbf{K}) \mathbf{C}_R) \right]^{-1}$$

while $(\delta(\mathbf{K}_d), \tilde{\delta}(\mathbf{K}_d))$ are the positive solutions of

$$\delta(\mathbf{K}_d) = \frac{1}{t} \mathrm{Tr} \mathbf{\Lambda} \left[ \sigma^2 (\mathbf{I} + \tilde{\delta}(\mathbf{K}_d) \mathbf{\Lambda}) \right]^{-1}$$

$$\tilde{\delta}(\mathbf{K}_d) = \frac{1}{t} \mathrm{Tr} \mathbf{C}_R \left[ \sigma^2 (\mathbf{I} + \delta(\mathbf{K}_d) \mathbf{C}_R) \right]^{-1}$$

As $\mathbf{K}^H \mathbf{C}_T \mathbf{K} = \mathbf{W} \mathbf{\Lambda} \mathbf{W}^H$, for each $\tilde{\kappa} > 0$, we have

$$\frac{1}{t} \mathrm{Tr} \mathbf{K}^H \mathbf{C}_T \mathbf{K} \left[ \sigma^2 (\mathbf{I} + \tilde{\kappa} \mathbf{K}^H \mathbf{C}_T \mathbf{K}) \right]^{-1} = \frac{1}{t} \mathrm{Tr} \mathbf{\Lambda} \left[ \sigma^2 (\mathbf{I} + \tilde{\kappa} \mathbf{\Lambda}) \right]^{-1}$$

Therefore, $(\delta(\mathbf{K}), \tilde{\delta}(\mathbf{K}))$ and $(\delta(\mathbf{K}_d), \tilde{\delta}(\mathbf{K}_d))$ are positive solutions of the same system. The uniqueness of the solutions yields $(\delta(\mathbf{K}), \tilde{\delta}(\mathbf{K})) = (\delta(\mathbf{K}_d), \tilde{\delta}(\mathbf{K}_d))$. From this, it is easy to check that $(\gamma(\mathbf{K}), \tilde{\gamma}(\mathbf{K})) = (\gamma(\mathbf{K}_d), \tilde{\gamma}(\mathbf{K}_d))$. $\hat{I}_{mmse}(\mathbf{K})$ can thus be written as

$$\hat{I}_{mmse}(\mathbf{K}) = \sum_{j=1}^{t} \log \left( \frac{1}{(\mathbf{I} + \tilde{\delta}(\mathbf{K}_d) \mathbf{W} \mathbf{\Lambda} \mathbf{W}^H)_{j,j}^{-1}} \right)$$

$(\mathbf{I} + \tilde{\delta}(\mathbf{K}_d) \mathbf{W} \mathbf{\Lambda} \mathbf{W}^H)_{j,j}^{-1}$ is given by

$$(\mathbf{I} + \tilde{\delta}(\mathbf{K}_d) \mathbf{W} \mathbf{\Lambda} \mathbf{W}^H)_{j,j}^{-1} = \sum_{k=1}^{t} \frac{|W_{j,k}|^2}{1 + \tilde{\delta}(\mathbf{K}_d) \lambda_k}$$

where $W_{j,k}$ is the entry $(j,k)$ of unitary matrix $\mathbf{W}$ and where $\mathbf{\Lambda} = \mathrm{Diag}(\lambda_1, \ldots, \lambda_t)$. The function $y \to \log \frac{1}{y}$ is convex on $\mathbb{R}^+$. As $\sum_{k=1}^{t} |W_{j,k}|^2 = 1$ (because $\mathbf{W}$ is unitary), we have

$$\log \left[ \frac{1}{\sum_{k=1}^{t} \frac{|W_{j,k}|^2}{1 + \tilde{\delta}(\mathbf{K}_d) \lambda_k}} \right] \leq \sum_{k=1}^{t} |W_{j,k}|^2 \log(1 + \tilde{\delta}(\mathbf{K}_d) \lambda_k)$$

Summing over $j$, and using that $\sum_{j=1}^{t} |W_{j,k}|^2 = 1$, we get that

$$\hat{I}_{mmse}(\mathbf{K}) \leq \sum_{k=1}^{t} \log(1 + \tilde{\delta}(\mathbf{K}_d)\lambda_k) = \hat{I}_{mmse}(\mathbf{K}_d)$$

We now establish that $\overline{J}_{mmse}(\mathbf{K}) \leq \overline{J}_{mmse}(\mathbf{K}_d)$.

We recall that

$$\sigma^2 \mathbf{T}_{T,j,j}(\mathbf{K}) = \sum_{k=1}^{t} \frac{|W_{j,k}|^2}{1 + \tilde{\delta}(\mathbf{K}_d)\lambda_k}$$

Similarly, we have

$$\left[(\sigma^2 \mathbf{T}_T)^2\right]_{j,j}(\mathbf{K}) = \sum_{k=1}^{t} \frac{|W_{j,k}|^2}{(1 + \tilde{\delta}(\mathbf{K}_d)\lambda_k)^2}$$

As $\sum_{k=1}^{t} |W_{j,k}|^2 = 1$, the convexity of function $x \to x^2$ implies that

$$\left[(\sigma^2 \mathbf{T}_T(\mathbf{K}))_{j,j}\right]^2 \leq \left[(\sigma^2 \mathbf{T}_T(\mathbf{K}))^2\right]_{j,j}$$

This implies that

$$\overline{J}_{mmse}(\mathbf{K}) \leq \frac{1}{t} \sum_{j=1}^{t} \left(1 - \sigma^2 \mathbf{T}_{T,j,j}(\mathbf{K})\right)^2$$

We also remark that

$$1 - \sigma^2 \mathbf{T}_{T,j,j}(\mathbf{K}) = \mathbf{w}_j (\mathbf{I} - \sigma^2 \mathbf{T}_T(\mathbf{K}_d)) \mathbf{w}_j^H$$

where $\mathbf{w}_j$ represents the row $j$ of $\mathbf{W}$. As matrix $\mathbf{I} - \sigma^2 \mathbf{T}(\mathbf{K}_d)$ is diagonal and matrix $\mathbf{W}$ is unitary, we have

$$\sum_{j=1}^{t} \left(\mathbf{w}_j(\mathbf{I} - \sigma^2 \mathbf{T}_T(\mathbf{K}_d))\mathbf{w}_j^H\right)^2 \leq \sum_{j=1}^{t} \left(1 - \sigma^2 \mathbf{T}_{T,j,j}(\mathbf{K}_d)\right)^2$$

We finally note that

$$\frac{1}{2\tilde{\delta}(\mathbf{K_d})^2} \frac{\tilde{\gamma}(\mathbf{K}_d)}{1 - \sigma^4 \gamma(\mathbf{K}_d)\tilde{\gamma}(\mathbf{K}_d)} \sum_{j=1}^{t} \left(1 - \sigma^2 \mathbf{T}_{T,j,j}(\mathbf{K}_d)\right)^2 = \overline{J}_{mmse}(\mathbf{K_d})$$

$\overline{J}_{mmse}(\mathbf{K}) \leq \overline{J}_{mmse}(\mathbf{K}_d)$ follows from the equalities $(\delta(\mathbf{K}), \tilde{\delta}(\mathbf{K})) = (\delta(\mathbf{K}_d), \tilde{\delta}(\mathbf{K}_d))$ and $(\gamma(\mathbf{K}), \tilde{\gamma}(\mathbf{K})) = (\gamma(\mathbf{K}_d), \tilde{\gamma}(\mathbf{K}_d))$. This completes the proof of Proposition 3.

Proposition 3 shows that there is no restriction to look an optimal precoder in the following set $\mathcal{K}_d$

$$\mathcal{K}_d = \{\mathbf{K} \in \mathcal{K}, \ \mathbf{K}^H \mathbf{C}_T \mathbf{K} \text{ diagonal}\} \tag{43}$$

This allows to formulate the evaluation of an optimal precoder as a $t$-dimensional optimization problem rather than a $t^2$ dimensional one. In order to state the corresponding result, we first slightly change our notations. If $\mathbf{K} \in \mathcal{K}_d$, the quantities $\delta(\mathbf{K}), \tilde{\delta}(\mathbf{K}), \ldots$ are actually functions of the entries of the diagonal matrix $\mathbf{\Lambda} = \mathbf{K}^H \mathbf{C}_T \mathbf{K}$. Therefore, for $\mathbf{K} \in \mathcal{K}_d$, $\delta(\mathbf{\Lambda}), \tilde{\delta}(\mathbf{\Lambda}), \ldots$ will be denoted $\delta(\mathbf{\Lambda}), \tilde{\delta}(\mathbf{\Lambda}), \ldots$.

The main result of this section is the following theorem.

*Theorem 2:* Let $\mathbf{C}_T = \mathbf{U}\mathbf{D}\mathbf{U}^H$ be the eigenvalues/eigenvectors decomposition of matrix $\mathbf{C}_T$. Let $\mathbf{\Lambda}_{opt} = \text{Diag}(\lambda_{1,opt}, \ldots, \lambda_{t,opt})$ be a positive diagonal matrix solution of the optimization problem

*Problem 1:* Maximize $\sum_{j=1}^{t} \log_2 \left(1 + \lambda_j \, \tilde{\delta}(\mathbf{\Lambda})\right) + \frac{1}{2} \frac{\sigma^4 \gamma(\mathbf{\Lambda}) \tilde{\gamma}(\mathbf{\Lambda})}{1 - \sigma^4 \gamma(\mathbf{\Lambda}) \tilde{\gamma}(\mathbf{\Lambda})}$ under the constraints

$$\mathbf{\Lambda} = \text{Diag}(\lambda_1, \ldots, \lambda_t) \geq 0, \ \frac{1}{t}\text{Tr}(\mathbf{D}^{-1}\mathbf{\Lambda}) \leq 1 \tag{44}$$

Then, matrix $\mathbf{K}_{opt}$ defined by

$$\mathbf{K}_{opt} = \mathbf{U}\mathbf{D}^{-1/2}\mathbf{\Lambda}_{opt}^{1/2} \tag{45}$$

belongs to $\mathcal{K}_d$, and maximizes $\overline{I}_{mmse}$.

**Proof.** In order to prove Theorem 2, we consider a precoding matrix $\mathbf{K} \in \mathcal{K}_d$, and denote $\mathbf{\Lambda} = \text{Diag}(\lambda_1, \lambda_2, \ldots, \lambda_t)$ the diagonal matrix $\mathbf{K}^H \mathbf{C}_T \mathbf{K}$. We assume that the diagonal entries $(\lambda_j)_{j=1,\ldots,t}$ of $\mathbf{\Lambda}$ are arranged in decreasing order. It is clear that $\mathbf{K}$ can be written as $\mathbf{K} = \mathbf{U}\mathbf{D}^{-1/2}\mathbf{\Theta}\mathbf{\Lambda}^{1/2}$ where $\mathbf{\Theta}$ is a unitary matrix. As $\frac{1}{t}\text{Tr}(\mathbf{K}\mathbf{K}^H)$ is supposed less than or equal to 1, matrices $\mathbf{\Lambda}$ and $\mathbf{\Theta}$ satisfy $\frac{1}{t}\text{Tr}\mathbf{D}^{-1}\mathbf{\Theta}\mathbf{\Lambda}\mathbf{\Theta}^H \leq 1$. Each precoder $\mathbf{K} \in \mathcal{K}_d$ can thus be parameterized by the unitary matrix $\mathbf{\Theta}$ and the positive diagonal matrix $\mathbf{\Lambda}$. As $\mathbf{K} \in \mathcal{K}_d$, one can check easily that $\overline{J}_{mmse}(\mathbf{\Lambda})$ reduces to $\frac{1}{2} \frac{\gamma(\mathbf{\Lambda})\tilde{\gamma}(\mathbf{\Lambda})}{1 - \sigma^4 \gamma(\mathbf{\Lambda})\tilde{\gamma}(\mathbf{\Lambda})}$. Therefore, Problem 1 is equivalent to the Problem

*Problem 2:* Maximize over $\mathbf{\Lambda}$ and $\mathbf{\Theta}$ $\sum_{j=1}^{t} \log_2 \left(1 + \lambda_j \, \tilde{\delta}(\mathbf{\Lambda})\right) + \frac{1}{2} \frac{\gamma(\mathbf{\Lambda})\tilde{\gamma}(\mathbf{\Lambda})}{1 - \sigma^4 \gamma(\mathbf{\Lambda})\tilde{\gamma}(\mathbf{\Lambda})}$ under the constraints

$$\mathbf{\Lambda} = \text{Diag}(\lambda_1, \ldots, \lambda_t) \geq 0, \ \mathbf{\Theta} \text{ unitary}, \ \frac{1}{t}\text{Tr}(\mathbf{D}^{-1}\mathbf{\Theta}\mathbf{\Lambda}\mathbf{\Theta}^H) \leq 1 \tag{46}$$

Let $(\mathbf{\Lambda}_*, \mathbf{\Theta}_*)$ be a solution of the above problem. The diagonal elements of $\mathbf{\Lambda}_*$ and $\mathbf{D}$ are arranged in decreasing order. Therefore (see the Appendix of [7]), the following inequality holds

$$\frac{1}{t}\text{Tr}\mathbf{D}^{-1}\mathbf{\Theta}_*\mathbf{\Lambda}_*\mathbf{\Theta}_*^H \geq \frac{1}{t}\text{Tr}\mathbf{D}^{-1}\mathbf{\Lambda}_* \tag{47}$$

Inequality (47) implies that if $(\mathbf{\Lambda}_*, \mathbf{\Theta}_*)$ is a solution of Problem 2, then, $(\mathbf{\Lambda}_*, \mathbf{I})$ is a solution of Problem 1. This shows that the optimization of $\overline{I}_{mmse}$ is equivalent to Problem 1. This completes the proof of Theorem 2.

*Remark 1:* We mention that it is not obvious that the singular vectors of the precoders that optimize the true mutual information $I_{mmse}$ have the structure (45). To our best knowledge, this is still an open question. One of the merit of the present asymptotic analysis is thus to show that the use of precoders (45) is relevant.

# V. Maximization of $\overline{I}_{mmse}$.

## A. Maximization of $\hat{I}_{mmse}$ in the case of i.i.d. channels.

Problem 1 cannot in general be solved in closed form. In this paragraph, we consider the case $r = t, \mathbf{C}_R = \mathbf{C}_T = \mathbf{I}_t$ for which some analytical results can be obtained. We establish in particular that the transmission scheme maximizing $\hat{I}_{mmse}$ is not a uniform power allocation along all the antennas, but an antenna selection scheme. This tends to indicate that in the context of i.i.d. channels, an antenna selection may provide higher values of $I_{mmse}$ than a uniform power allocation over the $t$ available transmit antennas. Therefore, even in the simplest channels context, the maximization of $I_{mmse}$ and of the usual Shannon mutual information $I$ are different problems.

Theorem 2 implies that precoders $\mathbf{K}_{opt}$ maximizing $\hat{I}_{mmse}$ can be written as $\mathbf{K}_{opt} = \mathbf{\Lambda}_{opt}^{1/2}$ where $\mathbf{\Lambda}_{opt}$ is solution of the problem

*Problem 3:* Maximize $\sum_{j=1}^{t} \log\left(1 + \lambda_j \tilde{\delta}(\mathbf{\Lambda})\right)$ under the constraints $\mathbf{\Lambda} = \mathrm{Diag}(\lambda_1, \ldots, \lambda_t) \geq 0$ and $\frac{1}{t}\mathrm{Tr}(\mathbf{\Lambda}) \leq 1$ where $\tilde{\delta}(\mathbf{\Lambda})$ is the unique positive solution of the equation

$$\sigma^2 \tilde{\delta} + \frac{1}{t}\sum_{j=1}^{t} \frac{\lambda_j \tilde{\delta}}{1 + \lambda_j \tilde{\delta}} = 1 \qquad (48)$$

It is easy to check that $\tilde{\delta}(\mathbf{\Lambda})$ is the positive solution of (48) in the particular context considered here. This justifies the statement of Problem 3. The solution of this problem is given in the following Proposition.

*Proposition 4:* The diagonal entries of the optimal matrices $\mathbf{\Lambda}_{opt}$ are either 0, either equal to $\frac{t}{s}$ where $s \leq t$, the number of non zero entries of $\mathbf{\Lambda}_{opt}$, is the integer that maximizes

$$s \log\left[\frac{\frac{t}{s} - 1 + \sigma^2 + \sqrt{(t/s - 1 + \sigma^2)^2 + 4\sigma^2}}{2\sigma^2}\right] \qquad (49)$$

**Proof**. We first show that any optimal matrix $\mathbf{\Lambda}_{opt}$ solution of Problem 3 verifies $\frac{1}{t}\mathrm{Tr}(\mathbf{\Lambda}_{opt}) = 1$. For this, we consider a positive diagonal matrix $\mathbf{\Lambda}$ for which $\frac{1}{t}\mathrm{Tr}(\mathbf{\Lambda}) < 1$, and establish that if $\mathbf{\Gamma}$ is the positive diagonal matrix with normalized trace 1 defined by $\mathbf{\Gamma} = \frac{1}{\frac{1}{t}\mathrm{Tr}(\mathbf{\Lambda})}\mathbf{\Lambda}$ , then $\hat{I}_{mmse}(\mathbf{\Gamma}) > \hat{I}_{mmse}(\mathbf{\Lambda})$. For this, we show that function $\mu \to \mu\tilde{\delta}(\mu\mathbf{\Lambda})$ is strictly increasing on $\mathbb{R}^+$. We remark that $\tilde{\delta}(\mu\mathbf{\Lambda})$ is the unique positive solution of the equation

$$\sigma^2 \tilde{\delta} + \frac{1}{t}\sum_{j=1}^{t} \frac{\lambda_j \mu \tilde{\delta}}{1 + \lambda_j \mu \tilde{\delta}} = 1 \qquad (50)$$

or equivalently, that $\mu\tilde{\delta}(\mu\mathbf{\Lambda})$ is the unique solution of the equation $g_\mu(\tilde{\rho}) = 1$ where $g_\mu$ is defined by

$$g_\mu(\tilde{\rho}) = \frac{\sigma^2}{\mu}\tilde{\rho} + \frac{1}{t}\sum_{j=1}^{t} \frac{\lambda_j \tilde{\rho}}{1 + \lambda_j \tilde{\rho}}$$

For each $\mu$, function $\tilde{\rho} \rightarrow g_\mu(\tilde{\rho})$ is strictly increasing. Moreover, if $\mu_1 < \mu_2$, then $g_{\mu_1}(\tilde{\rho}) > g_{\mu_2}(\tilde{\rho})$. From this, we get immediately that $\mu_1\tilde{\delta}(\mu_1\boldsymbol{\Lambda}) < \mu_2\tilde{\delta}(\mu_2\boldsymbol{\Lambda})$. We have thus shown that $\mu \rightarrow \mu\tilde{\delta}(\mu\boldsymbol{\Lambda})$ is strictly increasing. We put $\mu = \frac{1}{\frac{1}{t}\mathrm{Tr}(\boldsymbol{\Lambda})}$. As $\frac{1}{t}\mathrm{Tr}(\boldsymbol{\Lambda}) < 1$, $\mu$ is strictly greater than 1. Therefore, $\mu\tilde{\delta}(\mu\boldsymbol{\Lambda}) > \tilde{\delta}(\boldsymbol{\Lambda})$ or $\mu\tilde{\delta}(\boldsymbol{\Gamma}) > \tilde{\delta}(\boldsymbol{\Lambda})$. Hence,

$$\sum_{j=1}^{t} \log\left(1 + \mu\,\lambda_j\tilde{\delta}(\boldsymbol{\Gamma})\right) > \sum_{j=1}^{t} \log\left(1 + \lambda_j\tilde{\delta}(\boldsymbol{\Lambda})\right)$$

As the $(\mu\lambda_j)_{j=1,\ldots,t}$ coincide with the diagonal entries of matrix $\boldsymbol{\Gamma}$, the above inequality implies that $\hat{I}_{mmse}(\boldsymbol{\Gamma}) > \hat{I}_{mmse}(\boldsymbol{\Lambda})$.

The above discussion shows that the constraint $\frac{1}{t}\mathrm{Tr}(\boldsymbol{\Lambda}) \leq 1$ can be replaced by $\frac{1}{t}\mathrm{Tr}(\boldsymbol{\Lambda}) = 1$ in the statement of Problem 3. In order to characterize the solutions of the maximization problem, we replace the variables $(\lambda_j)_{j=1,\ldots,t}$ by the variables $(x_j)_{j=1,\ldots,t}$ defined by

$$x_j = \lambda_j\tilde{\delta}(\boldsymbol{\Lambda}) \tag{51}$$

for $j = 1,\ldots,t$. We claim that the maximization of $\hat{I}_{mmse}$ over the constraints $\lambda_j \geq 0$ for $j = 1,\ldots,t$ and $\frac{1}{t}\mathrm{Tr}(\boldsymbol{\Lambda}) = 1$ is equivalent to the following problem

*Problem 4:* Maximize $\sum_{j=1}^{t} \log(1 + x_j)$ under the constraints $x_j \geq 0$ for each $j = 1,\ldots,t$, and

$$\sigma^2 \frac{1}{t}\sum_{j=1}^{t} x_j + \frac{1}{t}\sum_{j=1}^{t} \frac{x_j}{1 + x_j} = 1 \tag{52}$$

Indeed, let $(x_j)_{j=1,\ldots,t}$ be positive numbers satisfying (52), and consider $\tilde{\delta} = \frac{1}{t}\sum_{j=1}^{t} x_j$ and $\lambda_j = \frac{x_j}{\tilde{\delta}}$ for $j = 1,\ldots,t$. The matrix $\boldsymbol{\Lambda} = \mathrm{Diag}(\lambda_1,\ldots,\lambda_t)$ is positive and satisfies $\frac{1}{t}\mathrm{Tr}(\boldsymbol{\Lambda}) = 1$. Moreover, $\tilde{\delta}$ is solution of the equation (48), which implies that $\tilde{\delta} = \tilde{\delta}(\boldsymbol{\Lambda})$. Conversely, if $\boldsymbol{\Lambda} = \mathrm{Diag}(\lambda_1,\ldots,\lambda_t)$ is positive and satisfies $\frac{1}{t}\mathrm{Tr}(\boldsymbol{\Lambda}) = 1$, the $(x_j)_{j=1,\ldots,t}$ defined by (51) are positive and satisfy the constraint (52). The conclusion follows from the observation that $\sum_{j=1}^{t} \log\left(1 + \lambda_j\tilde{\delta}(\boldsymbol{\Lambda})\right) = \sum_{j=1}^{t} \log(1 + x_j)$.

The Karusch- Kuhn-Tucker (KKT) conditions provide necessary conditions for optimality. If $\mathbf{x} = (x_1,\ldots,x_t)^T$ is a solution of Problem 4, then, it exists $\mu$ for which

$$\begin{array}{rcl} \frac{\partial \mathcal{L}(\mathbf{x},\mu)}{\partial x_j} & = & 0 \text{ if } x_j > 0 \\ \frac{\partial \mathcal{L}(\mathbf{x},\mu)}{\partial x_j} & \leq & 0 \text{ if } x_j = 0 \end{array} \tag{53}$$

where $\mathcal{L}(\mathbf{x},\mu)$ is defined by

$$\mathcal{L}(\mathbf{x},\mu) = \sum_{j=1}^{t} \log(1 + x_j) - \frac{1}{\mu}\left[\sigma^2\sum_{j=1}^{t} x_j + \sum_{j=1}^{t} \frac{x_j}{1 + x_j}\right]$$

If $x_j > 0$, we obtain

$$\sigma^2(1 + x_j) + \frac{1}{1 + x_j} = \mu$$

 

and if $x_j = 0$, we have

$$\mu \leq 1 + \sigma^2 \tag{54}$$

If $s \leq t$ is the number of non zero $x_j$'s, we have also

$$\mu = \sigma^2 + \sigma^2 \frac{1}{s} \sum_{j=1}^{t} x_j + \frac{1}{s} \sum_{x_j > 0} \frac{1}{1 + x_j} \tag{55}$$

or

$$\mu = \sigma^2 + \sigma^2 \frac{1}{s} \sum_{j=1}^{t} x_j + \frac{1}{s} \left( \sum_{j=1}^{t} \frac{1}{1 + x_j} - (t - s) \right)$$

Using the identity $\frac{x_j}{1+x_j} = 1 - \frac{1}{1+x_j}$, we get that the constraint (52) can also be written as

$$\frac{1}{t} \sum_{j=1}^{t} \frac{1}{1 + x_j} = \sigma^2 \frac{1}{t} \sum_{j=1}^{t} x_j$$

$\mu$ is therefore given by

$$\mu = \sigma^2 + 2\sigma^2 \frac{t}{s} \left( \frac{1}{t} \sum_{j=1}^{t} x_j \right) + 1 - \frac{t}{s} \tag{56}$$

We also note that $x_j > 0$ is a solution of the equation

$$\sigma^2 x_j^2 + (2\sigma^2 - \mu) x_j + (1 + \sigma^2 - \mu) = 0 \tag{57}$$

If $\mu > 1 + \sigma^2$, (54) implies that $s = t$. Moreover, the equation (57) has a single strictly positive solution $y$. Therefore, $x_j = y$ for $j = 1, \ldots, t$. Using the correspondence (51) between $\mathbf{x}$ and $\mathbf{\Lambda}$, we obtain immediately that $\mathbf{\Lambda} = \mathbf{I}_t$.

We now consider the case $\mu \leq 1 + \sigma^2$. If $\mu < 2\sigma$, equation (57) has no real solution. Therefore, $\mu$ must be greater than $2\sigma$. The equation must have at least a positive solution. As $1 + \sigma^2 - \mu \geq 0$, this implies that $\mu > 2\sigma^2$. In sum, $\mu$ must be greater than $\max(2\sigma, 2\sigma^2)$, and the equation (57) has 2 positive solutions $y_1$ and $y_2$ given by

$$y_1 = \frac{\mu - 2\sigma^2 + \sqrt{\mu^2 - 4\sigma^2}}{2\sigma^2}$$

$$y_2 = \frac{\mu - 2\sigma^2 - \sqrt{\mu^2 - 4\sigma^2}}{2\sigma^2}$$

Therefore, each non zero $x_j$ can be equal to $y_1$ or to $y_2$. We denote $\#\{j, x_j = y_1\}$ as $\frac{s}{2} + u$ and $\#\{j, x_j = y_2\}$ as $\frac{s}{2} - u$ where $u$ is an integer if $s$ is even and $u$ is the sum of $1/2$ with an integer if $s$ is odd. Note that if $(x_j)_{j=1,\ldots,t}$ is a solution of Problem 4, $u$ must be positive because $y_1 > y_2$ and

$$\sum_{j=1}^{t} \log(1 + x_j) = \left( u + \frac{s}{2} \right) \log(1 + y_1) + \left( u - \frac{s}{2} \right) \log(1 + y_2)$$

$\frac{1}{t} \sum_{j=1}^{t} x_j$ is given by

$$\frac{1}{t} \sum_{j=1}^{t} x_j = \frac{s}{t} \frac{\mu - 2\sigma^2}{2\sigma^2} + \frac{u}{t} \frac{\sqrt{\mu^2 - 4\sigma^2}}{\sigma^2}$$

Plugging this into (56) and solving the equation w.r.t. $\mu$ yields to

$$\mu = \sqrt{4\sigma^2 + \left(\frac{s}{2u}\sigma^2 + \frac{t-s}{2u}\right)^2} \tag{58}$$

This allows to express $y_1$ and $y_2$ in terms of $t, s, u, \sigma^2$. After some calculations, we obtain that $\sum_{j=1}^{t} \log(1 + x_j) = (u + \frac{s}{2}) \log(1 + y_1) + (u - \frac{s}{2}) \log(1 + y_2)$ is given by

$$\sum_{j=1}^{t} \log(1 + x_j) = \frac{s}{2} \log \frac{1}{\sigma^2} + u \log \left[ \frac{\sqrt{1 + b^2 u^2} + 1}{\sqrt{1 + b^2 u^2} - 1} \right] \tag{59}$$

where $b^2$ is defined by

$$b^2 = \frac{16\sigma^2}{(t - s + \sigma^2 s)^2}$$

It is easily seen that the righthandside of (59), considered as a function of $u$, is increasing on $\mathbb{R}^+$. Therefore, it is maximum for $u = \frac{s}{2}$. This implies that $\#\{j, x_j = y_1\} = s$ and $\#\{j, x_j = y_2\} = 0$. Moreover, the righthandside of (59) for $u = s/2$ coincides with (49). This completes the proof of Proposition 4.

We now check numerically that for certain values of $\sigma^2$, $s$ does not coincide with $t$. In figure 3, we have considered the case $r = t = 8$, and have represented the values of $\hat{I}_{mmse}$ for $s = 6$ and $s = 8$. It is clear that if the SNR is greater than 8 dB, then $s = 6$ provides higher values of $\hat{I}_{mmse}$. The values of $\overline{I}_{mmse}$ and $I_{mmse}$ are still higher for $s = 6$ rather than for $s = 8$. This confirms that the antenna selection scheme may be better than the uniform power allocation across all the transmit antennas. Figure 4 represents $I_{mmse}, \hat{I}_{mmse}, \overline{I}_{mmse}$ versus $s$ when the SNR is equal to $15dB$, and demonstrates that $s = 6$ is the optimum value of $I_{mmse}$.

We note that if $s \neq t$, function $\hat{I}_{mmse}$ reaches its maximum at different points because there are more than one diagonal matrix whose entries are either 0 either $\frac{t}{s}$. Function $\hat{I}_{mmse}$ is thus a non concave function of the precoding matrix. Using the trick introduced in [5], it is possible to establish that $I_{mmse}$ is itself, in general, non concave.

### B. Study of Problem 1.

We consider again the optimization of $\overline{I}_{mmse}$ in the bi-correlated case. Theorem 2 shows that the determination of an optimal precoder $\mathbf{K}_{opt}$ needs to solve the optimization Problem 1. As this problem cannot be solved in closed form, we use a gradient algorithm. We propose to parameterize $\lambda_j$ by $\lambda_j = \alpha_j^2$ in order to get rid of the constraint $\lambda_j \geq 0$, and to use a standard gradient algorithm with projection on the constraint $\frac{1}{t} \sum_{j=1}^{t} \frac{\alpha_j^2}{d_j} \leq 1$ at each iteration. Note that the convergence of this algorithm towards a global maximum of $\overline{I}_{mmse}$ is not guaranteed because this last function is probably non concave in general.

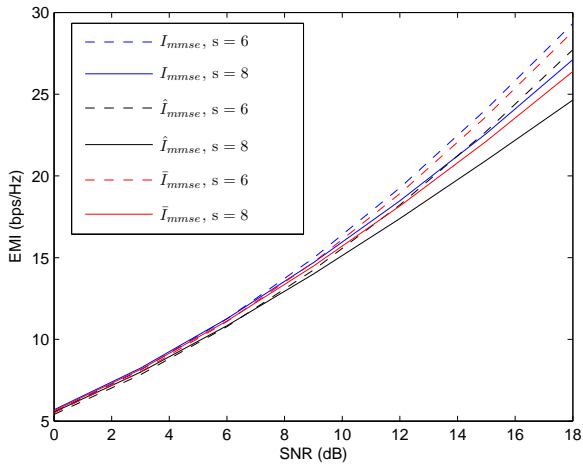

Fig. 3. Relevance of the antenna selection scheme, $s = 6$ versus $s = 8$

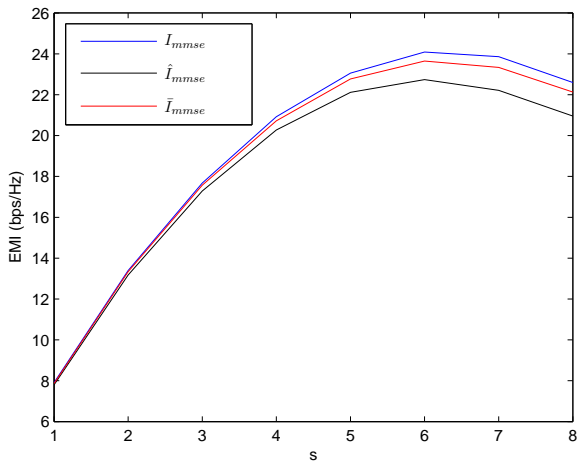

Fig. 4. Relevance of the antenna selection scheme, SNR = 15 dB

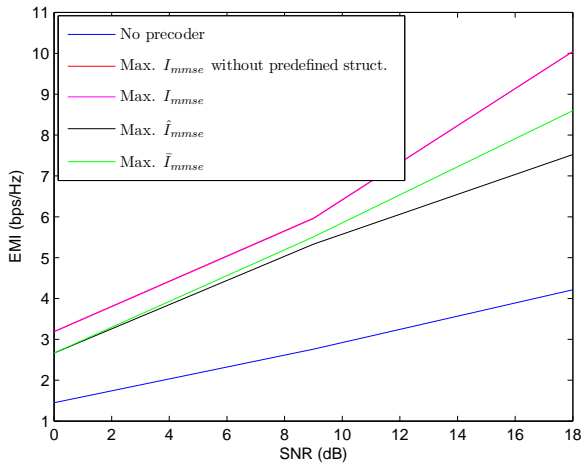

Fig. 5. Impact of precoding scheme

### C. Numerical illustration

We present some simulation results to illustrate the impact of the precoder optimization scheme in the case $r = t = 4$. $\mathbf{C}_T$ and $\mathbf{C}_R$ are generated according to model (38). In the present numerical experiment, $(\sigma_{\phi_T}, \phi_T) = (0.5, \frac{\pi}{4})$ and $(\sigma_{\phi_R}, \phi_R) = (0.4, \frac{\pi}{12})$.

In figure 5, we provide the mutual informations $I_{mmse}$ (evaluated using Monte Carlo simulations, 1000 channel realizations are used) corresponding to the following precoding schemes:

- (i) No precoding

- (ii) Maximization of $\overline{I}_{mmse}(\mathbf{K})$ for precoders structured as in (45)

- (iii) Maximization of $\hat{I}_{mmse}(\mathbf{K})$ for precoders structured as in (45)

- (iv) Maximization of $I_{mmse}(\mathbf{K})$ for precoders structured as in (45)

- (v) Maximization of $I_{mmse}(\mathbf{K})$ when the precoders have no particular structure.

The various maximizations are achieved by the gradient algorithm with projection on the relevant constraint. Note that the gradients of $I_{mmse}(\mathbf{K})$ w.r.t. the parameters $(\alpha_j)_{j=1,\ldots,t}$ and w.r.t. the entries of $\mathbf{K}$ have no closed form expression. At each iteration of the algorithm, they are evaluated by Monte Carlo simulations (1000 channel realizations are used). This explain why the direct maximization of $I_{mmse}$ leads to very high computational cost algorithms.

We now comment figure 5. We first compare precoding schemes (iv) and (v). The two curves match perfectly, showing that in practice, the structure (45) seems to optimize $I_{mmse}(\mathbf{K})$ even for $r = t = 4$. The comparison of schemes (ii) and (iii) indicates that the use of the improved approximation $\overline{I}_{mmse}$ allows to obtain significant gains for SNRs greater than 10 dB. We finally observe that the precoding schemes (ii) and (iv,v) provide very close mutual informations when SNR < 2 dB and SNR > 10 dB. Finally, the comparison of (i) (no precoding) with the other schemes shows that the precoding allows to increase significantly $I_{mmse}$.

We finally compare the processing time (on a $1.83GHz$ Intel) needed by schemes $(ii), (iii), (iv)$

| Precoding scheme | Processing time (s) |
|---|---|
| (ii) maximization of $\overline{I}_{mmse}$ | 0.39 |
| (iii) maximization of $\hat{I}_{mmse}$ | 0.25 |
| (iv) maximization of $I_{mmse}$ | 337.6 |

It is seen that the processing times needed to implement schemes (ii) and (iii) are almost 1000 times smaller than in the context of scheme (iv), while the use of the improved approximation $\overline{I}_{mmse}$ instead of $\hat{I}_{mmse}$ does not lead to a significant increase of the computational cost.

## VI. CONCLUDING REMARKS.

We summarize the advantages of our asymptotic analysis of $I_{mmse}$. It first allows to prove the relevance of precoders $\mathbf{K} = \mathbf{U}\mathbf{D}^{-1/2}\mathbf{\Lambda}^{1/2}$, where $\mathbf{\Lambda}$ is a positive diagonal matrix. Second, the entries of the optimum matrix $\mathbf{\Lambda}$ are solution of an optimization problem that can be solved by a computationally attractive gradient algorithm. If, in contrast, matrix $\mathbf{\Lambda}$ was designed to maximize the true mutual information $I_{mmse}$, the corresponding gradient algorithm would have a high computational cost. This is because this function of $\mathbf{\Lambda}$, as well as its derivatives w.r.t. the entries of $\mathbf{\Lambda}$, cannot be expressed in closed form. They have to be evaluated by Monte Carlo simulations, thus complicating a lot the maximization algorithm.

**Acknowledgements.** The authors thank Aris Moustakas for suggesting that the relative error of the approximation $\overline{I}_{mmse}$ was a $O(\frac{1}{t^2})$ term and not a $O(\frac{1}{t^{3/2}})$ term. Useful discussions with Walid Hachem and Jamal Najim are also acknowledged.

## APPENDIX A

### PROOF OF PROPOSITION 2.

The proof of Proposition 2 uses extensively the Nash-Poincaré inequality as well as an integration by part formula valid in the Gaussian random matrices context. The combined use of these two tools was introduced recently by Pastur in [14] in the context of simple models. This method was used in order to evaluate the asymptotic behaviour of the Shannon capacity of bi-correlated Rayleigh MIMO channels in [6] and of bi-correlated Rician MIMO channels in [5]. We however notice that Proposition 2 has not been established in [6] and [5].

Let $\Phi(\mathbf{Y})$ be a function of the entries of matrix $\mathbf{Y}$ defined by (9). Then, under certain extra assumptions on $\Phi$ (see [6]), the following Nash-Poincaré inequality holds true:

$$\mathrm{Var}\left(\Phi(\mathbf{Y})\right) \leq \frac{1}{t} \sum_{i=1}^{t} \sum_{j=1}^{r} d_i \tilde{d}_j \mathbb{E}\left[\left|\frac{\partial \Phi(\mathbf{Y})}{\partial Y_{i,j}}\right|^2 + \left|\frac{\partial \Phi(\mathbf{Y})}{\partial \overline{Y}_{i,j}}\right|^2\right] \ . \tag{60}$$

where $\overline{Y}_{i,j}$ represents the complex conjugate of $Y_{i,j}$. We also recall that the integration by part formula gives

$$\mathbb{E}\left[Y_{pq}\Phi(\mathbf{Y})\right] = \frac{d_p \tilde{d}_q}{t} \mathbb{E}\left[\frac{\partial \Phi(\mathbf{Y})}{\partial \overline{Y}_{pq}}\right] \ . \tag{61}$$

and

$$\mathbb{E}\left[\overline{Y}_{pq}\Phi(\mathbf{Y})\right] = \frac{d_p \tilde{d}_q}{t} \mathbb{E}\left[\frac{\partial \Phi(\mathbf{Y})}{\partial Y_{pq}}\right] \ . \tag{62}$$

We first establish (18). For this, we first introduce some notations. $\beta$ is defined by $\beta = \frac{1}{t}\mathrm{Tr}(\mathbf{D}\mathbf{Q})$ and $\alpha = \mathbb{E}(\beta)$. $\tilde{\mathbf{R}}$ is the $r \times r$ diagonal matrix given by

$$\tilde{\mathbf{R}} = \left[\sigma^2\left(\mathbf{I}_r + \alpha\tilde{\mathbf{D}}\right)\right]^{-1} \tag{63}$$

$\tilde{\alpha}$ is defined by $\tilde{\alpha} = \frac{1}{t}\mathrm{Tr}(\tilde{\mathbf{D}}\tilde{\mathbf{Q}})$, and $\mathbf{R}$ is the $t \times t$ diagonal matrix given by

$$\mathbf{R} = \left[\sigma^2\left(\mathbf{I}_t + \tilde{\alpha}\mathbf{D}\right)\right]^{-1} \tag{64}$$

If $x$ is a random variable, $\overset{\circ}{x}$ represents the random variable $\overset{\circ}{x} = x - \mathbb{E}(x)$.

Using calculations similar to [6], section 4.1, we obtain that

$$\mathbb{E}((\mathbf{Q}\mathbf{y}_j)_k \overline{Y}_{i,j}) = \frac{d_i}{t}\frac{\tilde{d}_j}{1 + \alpha\tilde{d}_j}\mathbb{E}(\mathbf{Q}_{k,i}) - \frac{\tilde{d}_j}{1 + \alpha\tilde{d}_j}\mathbb{E}(\overset{\circ}{\beta}(\mathbf{Q}\mathbf{y}_j)_k \overline{Y}_{i,j})$$

for each $k, i, j$. Summing over $j$ gives

$$\mathbb{E}\left((\mathbf{Q}\mathbf{Y}\mathbf{Y}^H)_{k,i}\right) = \sigma^2 d_i \tilde{\alpha}\mathbb{E}(\mathbf{Q}_{k,i}) - \sigma^2\mathbb{E}\left(\overset{\circ}{\beta}(\mathbf{Q}\mathbf{Y}\tilde{\mathbf{D}}\tilde{\mathbf{R}}\mathbf{Y}^H)_{k,i}\right) \tag{65}$$

Plugging the resolvent identity (see Eq. (10) of [6])

$$\mathbf{Q}_{k,i} = \frac{\delta(k-i)}{\sigma^2} - \frac{(\mathbf{Q}\mathbf{Y}\mathbf{Y}^H)_{k,i}}{\sigma^2} \tag{66}$$

into (65), we obtain

$$\mathbb{E}(\mathbf{Q}_{k,i}) = \frac{\delta(k-i)}{\sigma^2} - d_i\tilde{\alpha}\mathbb{E}(\mathbf{Q}_{k,i}) + \mathbb{E}\left(\overset{\circ}{\beta}(\mathbf{Q}\mathbf{Y}\tilde{\mathbf{D}}\tilde{\mathbf{R}}\mathbf{Y}^H)_{k,i}\right)$$

Solving w.r.t. $\mathbb{E}(\mathbf{Q}_{k,i})$, we get

$$\mathbb{E}(\mathbf{Q}_{k,i}) = R_{i,i}\delta(k-i) + \sigma^2\mathbb{E}\left(\overset{\circ}{\beta}(\mathbf{QY}\tilde{\mathbf{D}}\tilde{\mathbf{R}}\mathbf{Y}^H\mathbf{R})_{k,i}\right)$$

If $\mathbf{u}$ is a deterministic unit norm row vector, we eventually obtain

$$\mathbb{E}(\mathbf{u}\mathbf{Q}\mathbf{u}^H) = \mathbf{u}\mathbf{R}\mathbf{u}^H + \sigma^2\mathbb{E}\left(\overset{\circ}{\beta}\mathbf{u}\mathbf{Q}\mathbf{Y}\tilde{\mathbf{D}}\tilde{\mathbf{R}}\mathbf{Y}^H\mathbf{R}\mathbf{u}^H\right) \qquad (67)$$

We now prove that the second term of the righthandside of (67) can be bounded by a $O(\frac{1}{t^{3/2}})$ term independent of $\mathbf{u}$. As $\mathbb{E}(\overset{\circ}{\beta}) = 0$, the Schwartz inequality gives

$$\left|\mathbb{E}\left(\left(\overset{\circ}{\beta}\mathbf{u}\mathbf{Q}\mathbf{Y}\tilde{\mathbf{D}}\tilde{\mathbf{R}}\mathbf{Y}^H\mathbf{R}\mathbf{u}^H\right)\right| \le \left(\mathbb{E}\left|\overset{\circ}{\beta}\right|^2\right)^{1/2}\left[\text{Var}\left(\mathbf{u}\mathbf{Q}\mathbf{Y}\tilde{\mathbf{D}}\tilde{\mathbf{R}}\mathbf{Y}^H\mathbf{R}\mathbf{u}^H\right)\right]^{1/2} \qquad (68)$$

Using the first item of (15) in the case $\mathbf{M} = \mathbf{D}$, we get that $\left(\mathbb{E}|\overset{\circ}{\beta}|^2\right)^{1/2} = O(\frac{1}{t})$. In order to study the behaviour of the second term of the righthandside of (68), we establish the following lemma.

*Lemma 2:* Let $\mathbf{A}$ be a uniformly bounded diagonal deterministic matrix, $\mathbf{u}$ a unit norm deterministic row vector, and $\mathbf{v}$ a uniformly bounded deterministic row vector. Then,

$$\text{Var}(\mathbf{u}\mathbf{Q}\mathbf{Y}\mathbf{A}\mathbf{Y}^H\mathbf{v}^H) \le \frac{C}{t} \qquad (69)$$

where $C$ is a constant independent of $\mathbf{u}$, $\mathbf{v}$, and $\mathbf{A}$.

**Proof**. In order prove the lemma, we use the Nash-Poincaré inequality (60) in the case $\Phi(\mathbf{Y}) = \mathbf{u}\mathbf{Q}\mathbf{Y}\mathbf{A}\mathbf{Y}^H\mathbf{v}^H$. We define $\eta$ as $\eta = \mathbf{u}\mathbf{Q}\mathbf{Y}\mathbf{A}\mathbf{Y}^H\mathbf{v}^H$. Straightforward calculations lead to

$$\frac{\partial\eta}{\partial\overline{Y}_{i,j}} = -\mathbf{u}\mathbf{Q}\mathbf{y}_j(\mathbf{Q}\mathbf{Y}\mathbf{A}\mathbf{Y}^H\mathbf{v}^H)_i + A_{j,j}v_i\,\mathbf{u}\mathbf{Q}\mathbf{y}_j \qquad (70)$$

We establish that

$$\sum_{i,j}\mathbb{E}\left|\frac{\partial\eta}{\partial\overline{Y}_{i,j}}\right|^2 \le C \qquad (71)$$

where $C$ is a constant independent of $\mathbf{u}$, $\mathbf{v}$, $\mathbf{A}$. (70), $|A_{j,j}| \le \|\mathbf{A}\|$ and the Schwartz inequality imply that

$$\mathbb{E}\left|\frac{\partial\eta}{\partial\overline{Y}_{i,j}}\right|^2 \le 2|v_i|^2\|\mathbf{A}\|^2\mathbb{E}|\mathbf{u}\mathbf{Q}\mathbf{y}_j|^2 + 2\mathbb{E}\left(|(\mathbf{Q}\mathbf{Y}\mathbf{A}\mathbf{Y}^H\mathbf{v}^H)_i|^2|\mathbf{u}\mathbf{Q}\mathbf{y}_j|^2\right)$$

Summing over $i,j$ yields

$$\sum_{i,j}\mathbb{E}\left|\frac{\partial\eta}{\partial\overline{Y}_{i,j}}\right|^2 \le 2\|\mathbf{A}\|^2\|\mathbf{v}\|^2\mathbb{E}\|\mathbf{u}\mathbf{Q}\mathbf{Y}\|^2 + 2\mathbb{E}\left(\|\mathbf{Q}\mathbf{Y}\mathbf{A}\mathbf{Y}^H\mathbf{v}^H\|^2\|\mathbf{u}\mathbf{Q}\mathbf{Y}\|^2\right)$$

$\mathbb{E}(\|\mathbf{u}\mathbf{Q}\mathbf{Y}\|^2) = \mathbb{E}(\mathbf{u}\mathbf{Q}\mathbf{Y}\mathbf{Y}^H\mathbf{Q}\mathbf{u}^H)$. Using the resolvent identity (66), we obtain that $\mathbf{Q}\mathbf{Y}\mathbf{Y}^H = \mathbf{I} - \sigma^2\mathbf{Q}$. Therefore, $\mathbf{Q}\mathbf{Y}\mathbf{Y}^H\mathbf{Q} = \mathbf{Q} - \sigma^2\mathbf{Q}\mathbf{Q}$ and $\mathbf{Q}\mathbf{Y}\mathbf{Y}^H\mathbf{Q} \le \mathbf{Q}$. This implies that $\|\mathbf{u}\mathbf{Q}\mathbf{Y}\|^2 \le \mathbf{u}\mathbf{Q}\mathbf{u}^H$. As matrix $\mathbf{Q}$ satisfies $\mathbf{Q} \le \frac{\mathbf{I}}{\sigma^2}$, we obtain that

$$\|\mathbf{u}\mathbf{Q}\mathbf{Y}\|^2 \le \frac{1}{\sigma^2} \qquad (72)$$

In order to prove (71), it is thus sufficient to verify that $\mathbb{E}(\|\mathbf{QYAY}^H\mathbf{v}^H\|^2) \leq C$ where $C$ is a constant independent of $\mathbf{v}$ and $t$. For this, we remark that

$$\|\mathbf{QYAY}^H\mathbf{v}^H\|^2 \leq \frac{\mathbf{vYA}^H\mathbf{Y}^H\mathbf{YAY}^H\mathbf{v}^H}{\sigma^4}$$

A straightforward but tedious calculation gives

$$\mathbb{E}(\mathbf{vYA}^H\mathbf{Y}^H\mathbf{YAY}^H\mathbf{v}^H) = \mathbf{vD}^2\mathbf{v}^H\left|\frac{1}{t}\mathrm{Tr}(\mathbf{A}^H\tilde{\mathbf{D}})\right|^2 + \mathbf{vDv}^H\frac{1}{t}\mathrm{Tr}\mathbf{D}\frac{1}{t}\mathrm{Tr}(\mathbf{A}^H\tilde{\mathbf{D}}\mathbf{A}\tilde{\mathbf{D}})$$

As matrices $\mathbf{A}, \mathbf{D}, \tilde{\mathbf{D}}$ and vector $\mathbf{v}$ are uniformly bounded, we obtain that $\mathbb{E}(\|\mathbf{QYAY}^H\mathbf{u}^H\|^2) \leq C$. This, in turn, implies (71). One can show similarly that

$$\sum_{i,j}\mathbb{E}\left|\frac{\partial\eta}{\partial Y_{i,j}}\right|^2 \leq C$$

As the $(d_i)_{i=1,\ldots,t}$ and the $(\tilde{d}_j)_{j=1,\ldots,r}$ are uniformly bounded (see (14)), (60) provides immediately (69). Lemma 2 is thus established.

(67) and (68) imply that

$$\left|\mathbf{u}\left(\mathbb{E}(\mathbf{Q}) - \mathbf{R}\right)\mathbf{u}^H\right| \leq \frac{C}{t^{3/2}} \tag{73}$$

In order to complete the proof of (18), we use Theorem 3 of [6], and obtain that

$$\frac{1}{t}\mathrm{Tr}(\tilde{\mathbf{D}}\tilde{\mathbf{R}}) = \frac{1}{t}\mathrm{Tr}(\tilde{\mathbf{D}}\tilde{\mathbf{T}}) + O(\frac{1}{t^2})$$

or $\tilde{\alpha} = \tilde{\delta} + O(\frac{1}{t^2})$. It is easy to check that

$$|R_{i,i} - T_{i,i}| \leq \frac{d_{max}}{\sigma^2}|\tilde{\alpha} - \tilde{\delta}|$$

Therefore, $\max_i |R_{i,i} - T_{i,i}| \leq \frac{C}{t^2}$, and $|\mathbf{u}(\mathbf{R} - \mathbf{T})\mathbf{u}^H| \leq \frac{C}{t^2}$. Using (73), we eventually get (18).

We now establish (20). For this, we first prove the following lemma.

*Lemma 3:*

$$\mathbb{E}(\mathbf{Q}_{k,i}\overset{\circ}{\tilde{\mathbf{Q}}}_{k',i'}) = \frac{1}{t}\frac{1}{1 + \tilde{\alpha}d_i}\mathbb{E}\left[(\mathbf{QDQ})_{k,i'}(\mathbf{QY}\tilde{\mathbf{D}}\tilde{\mathbf{R}}\mathbf{Y}^H)_{k',i}\right] + \mathbb{E}\left[\overset{\circ}{\tilde{\beta}}(\mathbf{QY}\tilde{\mathbf{D}}\tilde{\mathbf{R}}\mathbf{Y}^H)_{k,i}\frac{1}{1 + \tilde{\alpha}d_i}\overset{\circ}{\tilde{\mathbf{Q}}}_{k',i'}\right] \tag{74}$$

**Proof.** We first note that (66) yields

$$\mathbb{E}(\mathbf{Q}_{k,i}\overset{\circ}{\tilde{\mathbf{Q}}}_{k',i'}) = -\frac{1}{\sigma^2}\mathbb{E}\left((\mathbf{QYY}^H)_{k,i}\overset{\circ}{\tilde{\mathbf{Q}}}_{k',i'}\right) \tag{75}$$

In order to be able to express $\mathbb{E}\left((\mathbf{QYY}^H)_{k,i}\overset{\circ}{\mathbf{Q}}_{k',i'}\right)$, we evaluate

$$\mathbb{E}\left((\mathbf{Qy}_j)_k\overline{Y}_{i,j}\overset{\circ}{\mathbf{Q}}_{k',i'}\right) = \sum_{p=1}^{t}\mathbb{E}(\mathbf{Q}_{k,p}Y_{p,j}\overline{Y}_{i,j}\overset{\circ}{\mathbf{Q}}_{k',i'})$$

For this, we use the integration by parts formula (61) in the case $\Phi(\mathbf{Y}) = \mathbf{Q}_{k,p}\overline{Y}_{i,j}\overset{\circ}{\mathbf{Q}}_{k',i'}$, and obtain

$$\mathbb{E}(\mathbf{Q}_{k,p}Y_{p,j}\overline{Y}_{i,j}\overset{\circ}{\mathbf{Q}}_{k',i'}) = \delta(p-i)\frac{d_p\tilde{d}_j}{t}\mathbb{E}(\mathbf{Q}_{k,p}\overset{\circ}{\mathbf{Q}}_{k',i'}) -$$
$$\frac{d_p\tilde{d}_j}{t}\mathbb{E}(\mathbf{Q}_{p,p}(\mathbf{Qy}_j)_k\overset{\circ}{\mathbf{Q}}_{k',i'}\overline{Y}_{i,j}) - \frac{d_p\tilde{d}_j}{t}\mathbb{E}(\mathbf{Q}_{k,p}\mathbf{Q}_{p,i'}(\mathbf{Qy}_j)_{k'}\overline{Y}_{i,j}) \quad (76)$$

Summing over $p$, and expressing $\beta = \frac{1}{t}\text{Tr}(\mathbf{DQ})$ as $\beta = \alpha + \overset{\circ}{\beta}$ provides

$$\mathbb{E}\left[(\mathbf{Qy}_j)_k\overline{Y}_{i,j}\overset{\circ}{\mathbf{Q}}_{k',i'}\right] = \frac{d_i\tilde{d}_j}{t}\mathbb{E}(\mathbf{Q}_{k,i}\overset{\circ}{\mathbf{Q}}_{k',i'}) - \frac{\tilde{d}_j}{t}\mathbb{E}\left[(\mathbf{QDQ})_{k,i'}(\mathbf{Qy}_j)_{k'}\overline{Y}_{i,j}\right] -$$
$$\alpha\tilde{d}_j\mathbb{E}\left[(\mathbf{Qy}_j)_k\overline{Y}_{i,j}\overset{\circ}{\mathbf{Q}}_{k',i'}\right] - \tilde{d}_j\mathbb{E}\left[\overset{\circ}{\beta}(\mathbf{Qy}_j)_k\overline{Y}_{i,j}\overset{\circ}{\mathbf{Q}}_{k',i'}\right] \quad (77)$$

Solving w.r.t. $\mathbb{E}\left[(\mathbf{Qy}_j)_k\overline{Y}_{i,j}\overset{\circ}{\mathbf{Q}}_{k',i'}\right]$ and summing over $j$ gives

$$\mathbb{E}\left((\mathbf{QYY}^H)_{k,i}\overset{\circ}{\mathbf{Q}}_{k',i'}\right) = \sigma^2 d_i\mathbb{E}(\mathbf{Q}_{k,i}\overset{\circ}{\mathbf{Q}}_{k',i'}) - \qquad\qquad (78)$$
$$\frac{\sigma^2}{t}\mathbb{E}\left[(\mathbf{QDQ})_{k,i'}(\mathbf{QY}\tilde{\mathbf{D}}\tilde{\mathbf{R}}\mathbf{Y}^H)_{k',i}\right] - \sigma^2\mathbb{E}\left[\overset{\circ}{\beta}\overset{\circ}{\mathbf{Q}}_{k',i'}(\mathbf{QY}\tilde{\mathbf{D}}\tilde{\mathbf{R}}\mathbf{Y}^H)_{k,i}\right]$$

Plugging (75) into (78) and solving w.r.t. $\mathbb{E}(\mathbf{Q}_{k,i}\overset{\circ}{\mathbf{Q}}_{k',i'})$ gives (74).

We define $\eta$ by $\eta = \mathbf{uQu}^H$. (74) yields immediately

$$\mathbb{E}(\overset{\circ}{\eta})^2 = \mathbb{E}(\eta\overset{\circ}{\eta}) = \frac{\sigma^2}{t}\mathbb{E}\left(\mathbf{uQDQu}^H\,\mathbf{uQY}\tilde{\mathbf{D}}\tilde{\mathbf{R}}\mathbf{Y}^H\mathbf{Ru}^H\right) + \sigma^2\mathbb{E}\left[\overset{\circ}{\beta}\overset{\circ}{\eta}\,\mathbf{uQY}\tilde{\mathbf{D}}\tilde{\mathbf{R}}\mathbf{Y}^H\mathbf{Ru}^H\right] \quad (79)$$

We define $\rho_1$ and $\rho_2$ by $\rho_1 = \mathbf{uQDQu}^H$ and $\rho_2 = \mathbf{uQY}\tilde{\mathbf{D}}\tilde{\mathbf{R}}\mathbf{Y}^H\mathbf{Ru}^H$. The term $\mathbb{E}\left(\mathbf{uQDQu}^H\,\mathbf{uQY}\tilde{\mathbf{D}}\tilde{\mathbf{R}}\mathbf{Y}^H\mathbf{Ru}^H\right)$ is given by

$$\mathbb{E}\left(\mathbf{uQDQu}^H\,\mathbf{uQY}\tilde{\mathbf{D}}\tilde{\mathbf{R}}\mathbf{Y}^H\mathbf{Ru}^H\right) = \mathbb{E}(\rho_1)\mathbb{E}(\rho_2) + \mathbb{E}(\overset{\circ}{\rho}_1\overset{\circ}{\rho}_2)$$

In order to evaluate $\mathbb{E}(\rho_i)$, $i = 1, 2$, we state the following Lemma

*Lemma 4:*

$$\left|\mathbb{E}\left(\mathbf{QDQ}\right)_{k,i} - \frac{d_i T_{k,i}}{1 - \sigma^4\gamma\tilde{\gamma}}\right| < \frac{C}{t} \quad (80)$$

Let $\mathbf{A}$ be a uniformly bounded diagonal deterministic matrix. Then,

$$\left|\mathbb{E}(\mathbf{QYAY}^H)_{k,i} - \sigma^2 d_i\frac{1}{t}\text{Tr}(\mathbf{A}\tilde{\mathbf{T}}\tilde{\mathbf{D}})T_{k,i}\right| < \frac{C}{t} \quad (81)$$

The proof uses again the resolvent identity (66), the Nash-Poincaré inequality, the integration by parts formula, Theorem 3 in [6], and is omitted.

Using Lemma (4), we get that

$$\left| \sigma^2 \mathbb{E}(\rho_1)\mathbb{E}(\rho_2) - \frac{\sigma^4 \widetilde{\gamma}}{1 - \sigma^4 \gamma \widetilde{\gamma}} \left( \mathbf{u}\mathbf{T}^2\mathbf{D}\mathbf{u}^H \right)^2 \right| \leq \frac{C}{t}$$

We verify that $\mathbb{E}(\overset{\circ}{\rho}_1\overset{\circ}{\rho}_2)$ is a $O(\frac{1}{t^{1/2}})$ term. We first remark that, as $\rho_1 \leq \frac{d_{max}}{\sigma^4}$, then $|\overset{\circ}{\rho}_1| \leq 2\frac{d_{max}}{\sigma^4}$. The Schwartz inequality gives

$$|\mathbb{E}(\overset{\circ}{\rho}_1\overset{\circ}{\rho}_2)| \leq 2\frac{d_{max}}{\sigma^4} \left( \mathbb{E}|\overset{\circ}{\rho}_2|^2 \right)^{1/2} \leq \frac{C}{t^{1/2}}$$

by Lemma 2. Finally, we show that $\mathbb{E}(\overset{\circ}{\beta}\overset{\circ}{\eta}\rho_2)$ is a $O(\frac{1}{t^{3/2}})$ term. We express this term as

$$\mathbb{E}(\overset{\circ}{\beta}\overset{\circ}{\eta}\rho_2) = \mathbb{E}(\overset{\circ}{\beta}\overset{\circ}{\eta})\mathbb{E}(\rho_2) + \mathbb{E}(\overset{\circ}{\beta}\overset{\circ}{\eta}\overset{\circ}{\rho}_2)$$

Lemma 4 implies that $\mathbb{E}(\rho_2)$ is uniformly bounded, while the Schwartz inequality gives $\mathbb{E}(\overset{\circ}{\beta}\overset{\circ}{\eta}) = O(\frac{1}{t^{3/2}})$. In order to evaluate $\mathbb{E}(\overset{\circ}{\beta}\overset{\circ}{\eta}\overset{\circ}{\rho}_2)$, we write

$$\mathbb{E}(\overset{\circ}{\beta}\overset{\circ}{\eta}\overset{\circ}{\rho}_2) = \mathbb{E}(\overset{\circ}{\beta}\eta\overset{\circ}{\rho}_2) - \mathbb{E}(\eta)\mathbb{E}(\overset{\circ}{\beta}\overset{\circ}{\rho}_2)$$

As $\eta \leq \frac{d_{max}}{\sigma^2}$, the Schwartz inequality gives immediately that

$$\mathbb{E}(\overset{\circ}{\beta}\overset{\circ}{\eta}\overset{\circ}{\rho}_2) = O(\frac{1}{t^{3/2}}).$$

Putting all the pieces together completes the proof of (20).

In order to establish (21), we first need to prove the following lemma. This lemma will also be useful to establish Lemma 1 below.

*Lemma 5:* Let $\mathbf{M}$ be a uniformly bounded deterministic matrix. Then,

$$\mathbb{E}\left| \frac{1}{t}\mathrm{Tr}\mathbf{M}\left( \mathbf{Q} - \mathbb{E}(\mathbf{Q}) \right) \right|^4 = O(\frac{1}{t^4}) \tag{82}$$

Moreover,

$$\sup_{\mathbf{u}, \|u\|=1} \mathbb{E}\left| \mathbf{u}\left( \mathbf{Q} - \mathbb{E}(\mathbf{Q}) \right)\mathbf{u}^H \right|^8 \leq \frac{C}{t^4} \tag{83}$$

We denote by $\rho$ the random variable $\rho = \frac{1}{t}\mathrm{Tr}\mathbf{M}\left( \mathbf{Q} - \mathbb{E}(\mathbf{Q}) \right)$. $\mathbb{E}|\rho|^4$ can be written as

$$\mathbb{E}|\rho|^4 = (\mathbb{E}|\rho|^2)^2 + \mathrm{Var}(\rho^2)$$

The first item of (15) implies that $(\mathbb{E}|\rho|^2)^2 = O(\frac{1}{t^4})$. In order to evaluate $\mathrm{Var}(\rho^2)$, we use the Nash-Poincaré inequality in the case $\Phi(\mathbf{Y}) = \rho^2$.

$$
\begin{aligned}
\frac{\partial \rho^2}{\partial \overline{Y}_{i,j}} &= 2\rho \frac{1}{t} \sum_{p,q} \frac{\partial \mathbf{Q}_{p,q}}{\partial \overline{Y}_{i,j}} \mathbf{M}_{q,p} \\
&= -2\rho \frac{1}{t} \sum_{p,q} \mathbf{Q}_{i,q} (\mathbf{Q}\mathbf{y}_j)_p \mathbf{M}_{q,p} \\
&= -2\rho \frac{1}{t} \sum_p (\mathbf{Q}\mathbf{M})_{i,p} (\mathbf{Q}\mathbf{y}_j)_p \\
&= -2\rho \frac{1}{t} (\mathbf{Q}\mathbf{M}\mathbf{Q}\mathbf{y}_j)_i
\end{aligned}
$$

Therefore,

$$
\sum_{i,j} \mathbb{E} \left| \frac{\partial \rho^2}{\partial \overline{Y}_{i,j}} \right|^2 = \frac{4}{t^2} \mathbb{E} \left( |\rho|^2 \mathrm{Tr}(\mathbf{Y}^H \mathbf{Q}\mathbf{M}^H \mathbf{Q}^2 \mathbf{M}\mathbf{Q}\mathbf{Y}) \right)
$$

Matrix $\mathbf{M}^H \mathbf{Q}^2 \mathbf{M}$ is uniformly bounded. Therefore,

$$
\mathrm{Tr}(\mathbf{Y}^H \mathbf{Q}\mathbf{M}^H \mathbf{Q}^2 \mathbf{M}\mathbf{Q}\mathbf{Y}) \leq C \, \mathrm{Tr}\mathbf{Y}^H \mathbf{Q}\mathbf{Q}\mathbf{Y} = C \, \mathrm{Tr}(\mathbf{Q}\mathbf{Q}\mathbf{Y}\mathbf{Y}^H)
$$

As $\mathbf{Q}\mathbf{Y}\mathbf{Y}^H = \mathbf{I} - \sigma^2 \mathbf{Q} \leq \mathbf{I}$, we obtain that

$$
\mathrm{Tr}(\mathbf{Y}^H \mathbf{Q}\mathbf{M}^H \mathbf{Q}^2 \mathbf{M}\mathbf{Q}\mathbf{Y}) \leq C \, \mathrm{Tr}(\mathbf{Q})
$$

Hence,

$$
\sum_{i,j} \mathbb{E} \left| \frac{\partial \rho^2}{\partial \overline{Y}_{i,j}} \right|^2 \leq \frac{C}{t} \mathbb{E} \left( |\rho|^2 \frac{1}{t} \mathrm{Tr}(\mathbf{Q}) \right) \leq \frac{C}{t} \mathbb{E}(|\rho|^2)
$$

As $\mathbb{E}(|\rho|^2) = O(\frac{1}{t^2})$, this implies that

$$
\sum_{i,j} \mathbb{E} \left| \frac{\partial \rho^2}{\partial \overline{Y}_{i,j}} \right|^2 = O(\frac{1}{t^3})
$$

We obtain similarly that

$$
\sum_{i,j} \mathbb{E} \left| \frac{\partial \rho^2}{\partial Y_{i,j}} \right|^2 = O(\frac{1}{t^3})
$$

(82) follows immediately from the Nash-Poincaré identity.

In order to prove (83), we first establish that

$$
\sup_{u, \|u\|=1} \mathbb{E} \left| \mathbf{u} \left( \mathbf{Q} - \mathbb{E}(\mathbf{Q}) \right) \mathbf{u}^H \right|^4 \leq \frac{C}{t^2} \tag{84}
$$

and

$$
\sup_{u, \|u\|=1} \mathbb{E} \left| \mathbf{u} \left( \mathbf{Q} - \mathbb{E}(\mathbf{Q}) \right) \mathbf{u}^H \right|^6 \leq \frac{C}{t^3} \tag{85}
$$

We consider a deterministic unit norm row vector $\mathbf{u}$ and denote by $\eta$ the term $\eta = \mathbf{u} \left( \mathbf{Q} - \mathbb{E}(\mathbf{Q}) \right) \mathbf{u}^H$. $\mathbb{E}|\eta|^4 = \left( \mathbb{E}|\eta|^2 \right)^2 + \mathrm{Var}(\eta^2)$. (20) implies that $\left( \mathbb{E}|\eta|^2 \right)^2 \leq \frac{C}{t^2}$ where $C$ is a constant which does not depend on $t$ and $\mathbf{u}$. In order to evaluate the term $\mathrm{Var}(\eta^2)$, we use the Nash-Poincaré inequality in the case $\Phi(\mathbf{Y}) = \eta^2$.

$$\frac{\partial \eta^2}{\partial \overline{Y}_{i,j}} = -2\eta \, \mathbf{u}\mathbf{Q}\mathbf{y}_j \, (\mathbf{Q}\mathbf{u}^H)_i$$

Therefore,

$$\sum_{i,j} \mathbb{E} \left| \frac{\partial \eta^2}{\partial \overline{Y}_{i,j}} \right|^2 = 4\mathbb{E} \left( |\eta|^2 \, \mathbf{u}\mathbf{Q}\mathbf{Y}\mathbf{Y}^H\mathbf{Q}\mathbf{u}^H \, \mathbf{u}\mathbf{Q}^2\mathbf{u}^H \right)$$

(72), $\mathbf{Q} \leq \frac{\mathbf{I}}{\sigma^2}$, and $\mathbb{E}|\eta|^2 \leq \frac{C}{t}$ imply that

$$\sum_{i,j} \mathbb{E} \left| \frac{\partial \eta^2}{\partial \overline{Y}_{i,j}} \right|^2 \leq \frac{C}{t}$$

We obtain similarly that

$$\sum_{i,j} \mathbb{E} \left| \frac{\partial \eta^2}{\partial Y_{i,j}} \right|^2 \leq \frac{C}{t}$$

The Nash-Poincaré inequality eventually gives $\mathrm{Var}(\eta^2) \leq \frac{C}{t^2}$. We have therefore proved (84). In order to establish (85), we write $\mathbb{E}|\eta|^6 = \left( \mathbb{E}|\eta|^3 \right)^2 + \mathrm{Var}(\eta^3)$. The Holder inequality and (84) imply that $\left( \mathbb{E}|\eta|^3 \right)^2 \leq \frac{C}{t^3}$. The term $\mathrm{Var}(\eta^3)$ is also evaluated using the Nash-Poincaré inequality.

$$\frac{\partial \eta^3}{\partial \overline{Y}_{i,j}} = -3\eta^2 \mathbf{u}\mathbf{Q}\mathbf{y}_j (\mathbf{Q}\mathbf{u}^H)_i$$

and

$$\sum_{i,j} \mathbb{E} \left| \frac{\partial \eta^3}{\partial \overline{Y}_{i,j}} \right|^2 = 9\mathbb{E} \left( |\eta|^4 \, \mathbf{u}\mathbf{Q}\mathbf{Y}\mathbf{Y}^H\mathbf{Q}\mathbf{u}^H \, \mathbf{u}\mathbf{Q}^2\mathbf{u}^H \right)$$

As $\mathbf{u}\mathbf{Q}\mathbf{Y}\mathbf{Y}^H\mathbf{Q}\mathbf{u}^H\mathbf{u}$ and $\mathbf{u}\mathbf{Q}^2\mathbf{u}^H$ are uniformly bounded, (84) implies that

$$\sum_{i,j} \mathbb{E} \left| \frac{\partial \eta^3}{\partial \overline{Y}_{i,j}} \right|^2 \leq \frac{C}{t^2}$$

Similarly,

$$\sum_{i,j} \mathbb{E} \left| \frac{\partial \eta^3}{\partial Y_{i,j}} \right|^2 \leq \frac{C}{t^2}$$

(85) follows immediately from the Nash-Poincaré inequality.

Starting from $\mathbb{E}|\eta|^8 = \left( \mathbb{E}|\eta|^4 \right)^2 + \mathrm{Var}(\eta^4)$, (83) is proved similarly.

In order to establish (21), we introduce the following notations:

$$\rho_{1,k} = \mathbf{v}_k\mathbf{Q}\mathbf{D}\mathbf{Q}\mathbf{v}_k^H, \rho_{2,k} = \mathbf{v}_k\mathbf{Q}\mathbf{Y}\tilde{\mathbf{D}}\tilde{\mathbf{R}}\mathbf{Y}^H\mathbf{R}\mathbf{v}_k^H, \eta_k = \mathbf{v}_k\mathbf{Q}\mathbf{v}_k^H$$

Using (79) and Lemma 4, it is easy to check that

$$\sum_k \mathrm{Var}(\kappa_k \eta_k) - \frac{1}{t} \frac{\sigma^4 \tilde{\gamma}}{1 - \sigma^4 \gamma \tilde{\gamma}} \sum_k (\kappa_k \mathbf{v}_k \mathbf{T}^2 \mathbf{D} \mathbf{v}_k^H)^2 = \mathbb{E}\left(\overset{\circ}{\beta}\left(\sum_k \kappa_k \overset{\circ}{\eta}_k \rho_{2,k}\right)\right) + O(\frac{1}{t})$$

It therefore remains to show that

$$\mathbb{E}\left(\overset{\circ}{\beta}\left(\sum_k \kappa_k \overset{\circ}{\eta}_k \rho_{2,k}\right)\right) = O(\frac{1}{t}) \tag{86}$$

For this, we write $\rho_{2,k} = \overset{\circ}{\rho}_{2,k} + \mathbb{E}(\rho_{2,k})$. Therefore,

$$\mathbb{E}\left(\overset{\circ}{\beta}\left(\sum_k \kappa_k \overset{\circ}{\eta}_k \rho_{2,k}\right)\right) = \mathbb{E}\left[\overset{\circ}{\beta}\left(\sum_k \kappa_k \mathbb{E}(\rho_{2,k}) \overset{\circ}{\eta}_k\right)\right] + \mathbb{E}\left[\overset{\circ}{\beta}\left(\sum_k \kappa_k \overset{\circ}{\rho}_{2,k} \overset{\circ}{\eta}_k\right)\right] \tag{87}$$

The term $\mathbb{E}\left[\overset{\circ}{\beta}\left(\sum_k \kappa_k \mathbb{E}(\rho_{2,k}) \overset{\circ}{\eta}_k\right)\right]$ can also be written as $\mathbb{E}\left(\overset{\circ}{\beta} \mathrm{Tr}(\mathbf{M}\overset{\circ}{\mathbf{Q}})\right)$ where $\mathbf{M}$ is the deterministic matrix defined by

$$\mathbf{M} = \sum_k \kappa_k \mathbb{E}(\rho_{2,k}) \mathbf{v}_k^H \mathbf{v}_k$$

Lemma 4 implies that $\sup_k |\mathbb{E}(\rho_{2,k})| < C$. Therefore, matrix $\mathbf{M}$ is uniformly bounded. The first item of (15) thus implies that $\mathbb{E}\left|\mathrm{Tr}(\mathbf{M}\overset{\circ}{\mathbf{Q}})\right|^2 = O(1)$. Similarly, $\mathbb{E}|\overset{\circ}{\beta}|^2 = O(\frac{1}{t^2})$ holds. The Schwartz inequality shows that $\mathbb{E}\left(\overset{\circ}{\beta} \mathrm{Tr}(\mathbf{M}\overset{\circ}{\mathbf{Q}})\right) = O(\frac{1}{t})$.

In order to evaluate the second term of the righthandside of (87), we remark that

$$|\mathbb{E}(\overset{\circ}{\beta}\overset{\circ}{\eta}_k \overset{\circ}{\rho}_{2,k})| \leq (\mathbb{E}|\overset{\circ}{\rho}_{2,k}|^2)^{1/2} \, (\mathbb{E}|\overset{\circ}{\beta}_k|^4)^{1/4} \, (\mathbb{E}|\overset{\circ}{\eta}_k|^4)^{1/4}$$

Lemma 2 implies that $(\mathbb{E}|\overset{\circ}{\rho}_{2,k}|^2)^{1/2} = O(\frac{1}{t^{1/2}})$, and (84) gives $(\mathbb{E}|\overset{\circ}{\eta}_k|^4)^{1/4} = O(\frac{1}{t^{1/2}})$. As $(\mathbb{E}|\overset{\circ}{\beta}_k|^4)^{1/4} = O(\frac{1}{t})$ by (82), we get that

$$\sup_k |\mathbb{E}(\overset{\circ}{\beta}\overset{\circ}{\eta}_k \overset{\circ}{\rho}_{2,k})| = O(\frac{1}{t^2})$$

This, in turn, implies that the second term of the righthandside of (87) is a $O(\frac{1}{t})$ term. This completes the proof of (21).

We finally prove (19). We just sketch the proof because similar arguments have been used in order to establish Lemma 3. We evaluate $\mathbb{E}(\overset{\circ}{\mathbf{Q}}_{k_1,i_1} \overset{\circ}{\mathbf{Q}}_{k_2,i_2} \overset{\circ}{\mathbf{Q}}_{k_3,i_3})$ for each integers $(i_1, k_1, i_2, k_2, i_3, k_3)$. We first calculate $\mathbb{E}(\mathbf{Q}_{k_1,i_1} \overset{\circ}{\mathbf{Q}}_{k_2,i_2} \overset{\circ}{\mathbf{Q}}_{k_3,i_3})$. For this, we use the resolvent identity (66) and write

$$\mathbb{E}(\mathbf{Q}_{k_1,i_1} \overset{\circ}{\mathbf{Q}}_{k_2,i_2} \overset{\circ}{\mathbf{Q}}_{k_3,i_3}) = \frac{\delta(k_1 - i_1)}{\sigma^2} \mathbb{E}(\overset{\circ}{\mathbf{Q}}_{k_2,i_2} \overset{\circ}{\mathbf{Q}}_{k_3,i_3}) - \frac{1}{\sigma^2} \mathbb{E}\left[(\mathbf{Q}\mathbf{Y}\mathbf{Y}^H)_{k_1,i_1} \overset{\circ}{\mathbf{Q}}_{k_2,i_2} \overset{\circ}{\mathbf{Q}}_{k_3,i_3}\right]$$

Using the integration by parts formula as in the proof of Lemma 3, we obtain that

$$
\begin{aligned}
\tfrac{1}{\sigma^2}\mathbb{E}\left[(\mathbf{QYY}^H)_{k_1,i_1}\mathring{\mathbf{Q}}_{k_2,i_2}\mathring{\mathbf{Q}}_{k_3,i_3}\right] \;=\; & \tilde\alpha d_{i_1}\mathbb{E}(\mathbf{Q}_{k_1,i_1}\mathring{\mathbf{Q}}_{k_2,i_2}\mathring{\mathbf{Q}}_{k_3,i_3}) \\
& -\tfrac{1}{t}\mathbb{E}\left[(\mathbf{QDQ})_{k_1,i_2}(\mathbf{QY\tilde{D}\tilde{R}Y}^H)_{k_2,i_1}\mathring{\mathbf{Q}}_{k_3,i_3}\right] \\
& -\tfrac{1}{t}\mathbb{E}\left[(\mathbf{QDQ})_{k_1,i_3}(\mathbf{QY\tilde{D}\tilde{R}Y}^H)_{k_3,i_1}\mathring{\mathbf{Q}}_{k_2,i_2}\right] \\
& -\mathbb{E}\left[\mathring{\tilde\beta}(\mathbf{QY\tilde{D}\tilde{R}Y}^H)_{k_i,i_1}\mathring{\mathbf{Q}}_{k_2,i_2}\mathring{\mathbf{Q}}_{k_3,i_3}\right]
\end{aligned}
$$

Plugging (66) into the above equation and solving w.r.t. $\mathbb{E}(\mathbf{Q}_{k_1,i_1}\mathring{\mathbf{Q}}_{k_2,i_2}\mathring{\mathbf{Q}}_{k_3,i_3})$, we obtain that

$$
\begin{aligned}
\mathbb{E}(\mathbf{Q}_{k_1,i_1}\mathring{\mathbf{Q}}_{k_2,i_2}\mathring{\mathbf{Q}}_{k_3,i_3}) \;=\; & \mathbf{R}_{k_1,k_1}\delta(k_1-i_1)\mathbb{E}(\mathring{\mathbf{Q}}_{k_2,i_2}\mathring{\mathbf{Q}}_{k_3,i_3}) \\
& +\tfrac{\sigma^2}{t}\mathbb{E}\left[(\mathbf{QDQ})_{k_1,i_2}(\mathbf{QY\tilde{D}\tilde{R}Y}^H\mathbf{R})_{k_2,i_1}\mathring{\mathbf{Q}}_{k_3,i_3}\right] \\
& +\tfrac{\sigma^2}{t}\mathbb{E}\left[(\mathbf{QDQ})_{k_1,i_3}(\mathbf{QY\tilde{D}\tilde{R}Y}^H\mathbf{R})_{k_3,i_1}\mathring{\mathbf{Q}}_{k_2,i_2}\right] \\
& +\sigma^2\mathbb{E}\left[\mathring{\tilde\beta}(\mathbf{QY\tilde{D}\tilde{R}Y}^H\mathbf{R})_{k_i,i_1}\mathring{\mathbf{Q}}_{k_2,i_2}\mathring{\mathbf{Q}}_{k_3,i_3}\right]
\end{aligned}
$$

Writing $\mathbb{E}(\mathbf{Q}_{k_1,i_1}\mathring{\mathbf{Q}}_{k_2,i_2}\mathring{\mathbf{Q}}_{k_3,i_3})$ as

$$
\mathbb{E}(\mathbf{Q}_{k_1,i_1}\mathring{\mathbf{Q}}_{k_2,i_2}\mathring{\mathbf{Q}}_{k_3,i_3}) = \mathbb{E}(\mathbf{Q}_{k_1,i_1})\mathbb{E}(\mathring{\mathbf{Q}}_{k_2,i_2}\mathring{\mathbf{Q}}_{k_3,i_3}) + \mathbb{E}(\mathring{\mathbf{Q}}_{k_1,i_1}\mathring{\mathbf{Q}}_{k_2,i_2}\mathring{\mathbf{Q}}_{k_3,i_3})
$$

and using (73), we obtain that

$$
\begin{aligned}
\mathbb{E}(\mathring{\mathbf{Q}}_{k_1,i_1}\mathring{\mathbf{Q}}_{k_2,i_2}\mathring{\mathbf{Q}}_{k_3,i_3}) \;=\; & \tfrac{\sigma^2}{t}\mathbb{E}\left[(\mathbf{QDQ})_{k_1,i_2}(\mathbf{QY\tilde{D}\tilde{R}Y}^H\mathbf{R})_{k_2,i_1}\mathring{\mathbf{Q}}_{k_3,i_3}\right] \\
& +\tfrac{\sigma^2}{t}\mathbb{E}\left[(\mathbf{QDQ})_{k_1,i_3}(\mathbf{QY\tilde{D}\tilde{R}Y}^H\mathbf{R})_{k_3,i_1}\mathring{\mathbf{Q}}_{k_2,i_2}\right] \\
& +\sigma^2\mathbb{E}\left[\mathring{\tilde\beta}(\mathbf{QY\tilde{D}\tilde{R}Y}^H\mathbf{R})_{k_i,i_1}\mathring{\mathbf{Q}}_{k_2,i_2}\mathring{\mathbf{Q}}_{k_3,i_3}\right] + o(\tfrac{1}{t^2})
\end{aligned}
\tag{88}
$$

We consider a unit norm deterministic row vector $\mathbf{u}$ and define $\eta = \mathbf{uQu}, \rho_1 = \mathbf{uQDQu}^H$ and $\rho_2 = \mathbf{uQY\tilde{D}\tilde{R}Y}^H\mathbf{Ru}^H$. Using (88), we get that

$$
\mathbb{E}(\mathring{\eta}^3) = \frac{2\sigma^2}{t}\mathbb{E}(\rho_1\rho_2\mathring{\eta}) + \sigma^2\mathbb{E}(\mathring{\tilde\beta}\rho_2\mathring{\eta}^2)
$$

We write $\mathbb{E}(\rho_1\rho_2\mathring{\eta})$ as

$$
\mathbb{E}(\rho_1\rho_2\mathring{\eta}) = \mathbb{E}(\rho_1)\mathbb{E}(\mathring{\rho}_2\mathring{\eta}) + \mathbb{E}(\rho_2)\mathbb{E}(\mathring{\rho}_1\mathring{\eta}) + \mathbb{E}(\mathring{\rho}_1\mathring{\rho}_2\mathring{\eta})
$$

$\mathbb{E}(\rho_1)$ is uniformly bounded while $\mathbb{E}(\mathring{\rho}_2\mathring{\eta})$ is a $O(\tfrac{1}{t})$ term. $\mathbb{E}(\rho_2)\mathbb{E}(\mathring{\rho}_1\mathring{\eta})$ is a $O(\tfrac{1}{t})$ term for the same reasons. Finally, we remark that $|\mathring{\rho}_1| \leq \frac{2d_{max}}{\sigma^4}$. Therefore, $\mathbb{E}(\mathring{\rho}_1\mathring{\rho}_2\mathring{\eta})$ is a $O(\tfrac{1}{t})$ term, and $\frac{2\sigma^2}{t}\mathbb{E}(\rho_1\rho_2\mathring{\eta})$ is a $O(\tfrac{1}{t^2})$ term.

In order to evaluate $\mathbb{E}(\mathring{\beta}\rho_2\mathring{\eta}^2)$, we write

$$\mathbb{E}(\mathring{\beta}\rho_2\mathring{\eta}^2) = \mathbb{E}(\rho_2)\mathbb{E}(\mathring{\beta}\mathring{\eta}^2) + \mathbb{E}(\mathring{\beta}\mathring{\rho}_2\mathring{\eta}^2)$$

$\mathbb{E}(\rho_2)$ is uniformly bounded. $\mathbb{E}(\mathring{\beta}\mathring{\eta}^2) = O(\frac{1}{t^2})$ holds by the Schwartz inequality. We finally write that

$$|\mathbb{E}(\mathring{\beta}\mathring{\rho}_2\mathring{\eta}^2)| \leq \left(\mathbb{E}|\mathring{\rho}_2|^2\right)^{1/2} \left(\mathbb{E}|\mathring{\beta}|^4\right)^{1/4} \left(\mathbb{E}|\mathring{\eta}|^8\right)^{1/4}$$

and use Lemma 5 to justify that $\mathbb{E}(\mathring{\beta}\mathring{\rho}_2\mathring{\eta}^2) = O(\frac{1}{t^2})$. This completes the proof of (19).

## Appendix B

### Proof of Lemma 1.

We first establish that

$$\mathbb{E}\left(\log(1+\epsilon_j)\right)^2 < C \tag{89}$$

for some constant $C$ independent of $j$ and $t$. For this, we remark that

$$\sigma^2 \mathbf{Q}_{T,j,j} = \frac{1}{1+\beta_j} \tag{90}$$

is less than 1. Therefore, $-\mathbb{E}(\log(\sigma^2\mathbf{Q}_{T,j,j})) \geq 0$. As $\log(1+\epsilon_j)$ is equal to $\log(\sigma^2\mathbf{Q}_{T,j,j}) - \mathbb{E}(\log(\sigma^2\mathbf{Q}_{T,j,j}))$, we get that

$$\log(1+\epsilon_j) \geq \log(\sigma^2\mathbf{Q}_{T,j,j})) = -\log(1+\beta_j)$$

$\beta_j > 0$ implies that $\log(1+\beta_j) \leq \beta_j$. Therefore, $\log(1+\epsilon_j) \geq -\beta_j$ and $(\log(1+\epsilon_j))^2 \leq (\beta_j)^2$. In order to prove (89), it is thus sufficient to establish that $\mathbb{E}(\beta_j^2) \leq C$. We denote by $\mathbf{h}_j$ the column $j$ of matrix $\mathbf{H}$. $\beta_j$ is upperbounded by the match filter bound $\frac{\|\mathbf{h}_j\|^2}{\sigma^2}$. Using the expression of vector $\mathbf{h}_j$ in terms of matrices $\mathbf{C}_R$, $\mathbf{C}_T$ and $\mathbf{H}_{iid}$, it is easy to check that

$$\mathbb{E}(\|\mathbf{h}_j\|^4) \leq C$$

for some constant $C$ independent of $j$ and $t$. This completes the proof of (89). Note that (89) implies that for each $j$, $\mathbb{E}|\log(1+\epsilon_j)| < \infty$, a property which was implicitly assumed in the proof of Theorem 1.

We now complete the proof of Lemma 1. We consider a constant $A \in (0,1)$, and express $\log(1+\epsilon_j)$ as

$$\log(1+\epsilon_j) = \sum_{k=1}^{\infty}(-1)^{k-1}\frac{\epsilon_j^k}{k}\mathbb{I}_{|\epsilon_j|<A} + \log(1+\epsilon_j)\mathbb{I}_{|\epsilon_j|\geq A}$$

where for any set $\mathcal{B}$, $\mathbb{I}_{\mathcal{B}}$ is equal to 1 on $\mathcal{B}$ and 0 outside $\mathcal{B}$. This leads to the following expression of $\mathbb{E}(r_j)$

$$\mathbb{E}(r_j) = \sum_{k=4}^{\infty} (-1)^{k-1} \mathbb{E}\left(\frac{\epsilon_j^k}{k}\mathbb{I}_{|\epsilon_j|<A}\right) - \mathbb{E}\left(\epsilon_j\mathbb{I}_{|\epsilon_j|\geq A}\right) + \frac{1}{2}\mathbb{E}\left(\epsilon_j^2\mathbb{I}_{|\epsilon_j|\geq A}\right) - \frac{1}{3}\mathbb{E}\left(\epsilon_j^3\mathbb{I}_{|\epsilon_j|\geq A}\right) + \mathbb{E}\left(\log(1+\epsilon_j)\mathbb{I}_{|\epsilon_j|\geq A}\right)$$

(91)

Using (28) and (83), we remark that

$$\mathbb{E}|\epsilon_j|^8 \leq \frac{C}{t^4} \tag{92}$$

From the Markov inequality and the Holder inequality, we obtain that

$$P(|\epsilon_j| > A) \leq \frac{C}{t^4} \tag{93}$$

and

$$\begin{aligned}
\mathbb{E}\left((|\epsilon_j|^6\right) &\leq \frac{C}{t^3} \\
\mathbb{E}\left((|\epsilon_j|^4\right) &\leq \frac{C}{t^2} \\
\mathbb{E}\left(|\epsilon_j|^3\right) &\leq \frac{C}{t^{3/2}} \\
\mathbb{E}\left(|\epsilon_j|^2\right) &\leq \frac{C}{t}
\end{aligned} \tag{94}$$

By the Schwartz inequality,

$$\left|\mathbb{E}\left(\epsilon_j\mathbb{I}_{|\epsilon_j|\geq A}\right)\right| \leq \left(P(|\epsilon_j| > A)\right)^{1/2}\left(\mathbb{E}|\epsilon_j|^2\right)^{1/2}$$

(93) and (94) thus imply that $\left|\mathbb{E}\left(\epsilon_j\mathbb{I}_{|\epsilon_j|\geq A}\right)\right|$ is upperbounded by $\frac{C}{t^{5/2}}$. We obtain similarly that

$$\mathbb{E}\left(\epsilon_j^2\mathbb{I}_{|\epsilon_j|\geq A}\right) \leq \frac{C}{t^3}$$

and

$$\mathbb{E}\left(\epsilon_j^3\mathbb{I}_{|\epsilon_j|\geq A}\right) \leq \frac{C}{t^{7/2}}$$

Using (89), (93) and the Schwartz inequality yields

$$\left|\mathbb{E}\left(\log(1+\epsilon_j)\mathbb{I}_{|\epsilon_j|\geq A}\right)\right| \leq \frac{C}{t^2}$$

We now establish that

$$\sum_{k=4}^{\infty} \mathbb{E}\left(\frac{|\epsilon_j|^k}{k}\mathbb{I}_{|\epsilon_j|<A}\right) \leq \frac{C}{t^2} \tag{95}$$

For $k \geq 4$, $\mathbb{E}\left(|\epsilon_j|^k\mathbb{I}_{|\epsilon_j|<A}\right) \leq A^{k-4}\mathbb{E}(|\epsilon_j|^4)$. Therefore,

$$\sum_{k=4}^{\infty} \mathbb{E}\left(\frac{|\epsilon_j|^k}{k}\mathbb{I}_{|\epsilon_j|<A}\right) \leq \left(\sum_{k=0}^{\infty}\frac{A^k}{k+4}\right)\mathbb{E}(|\epsilon_j|^4)$$

As $A < 1$, $\sum_{k=0}^{\infty}\frac{A^k}{k+4} < \infty$ so that (95) follows from (94). Putting all the pieces together gives $|\mathbb{E}(r_j)| \leq \frac{C}{t^2}$.